\documentclass[12pt]{article}
\usepackage{amsmath, amsthm}
\usepackage{multirow}
\usepackage{amsfonts}
\usepackage{amssymb,graphics,psfrag}
\usepackage{array,epsfig,multirow,stmaryrd,graphicx,setspace}
\usepackage{comment}
\usepackage[scale=0.73]{geometry}

\numberwithin{equation}{section}
\numberwithin{table}{section}

  \let\Th=\Theta \let\S=\Sigma
\def\2{{1\over2}}
\let\<=\langle \let\>=\rangle
\def\new#1\endnew{{\bf #1}}
\def\ifundefined#1{\expandafter\ifx\csname#1\endcsname\relax}
\ifundefined{draftmode}\else    \input draftmode        \fi
\let\Msize=\footnotesize             
\def\BM{\Msize\begin{matrix}}           \def\EM{\end{matrix}}
\def\MN M:#1 #2 N:#3 #4 {{(#1_{#2},#3_{#4})}}
\def\MNH M:#1 #2 N:#3 #4 H:#5,#6 [#7]{{(#1_{#2},#3_{#4})^{#5,#6}_{#7}}}

\newcommand{\tr}{{\rm Tr}}

\def\CE{{\cal E}}

\def\CH{{\cal H}}

\def\CK{{\cal K}}
\def\CL{{\cal L}}

\def\CN{{\cal N}}
\def\CO{{\cal O}}

\def\CU{{\cal U}}

\def\CW{{\cal W}}

\newcommand{\be}{\begin{equation}}
\newcommand{\ee}{\end{equation}}
\newcommand{\bea}{\begin{eqnarray}}
\newcommand{\eea}{\end{eqnarray}}

\def\IZ{{\mathbb Z}}
\def\IR{{\mathbb R}}
\def\IC{{\mathbb C}}
\def\IP{{\mathbb P}}
\def\IT{{\mathbb T}}
\def\IS{{\mathbb S}}

\newcommand{\re}{{\rm e}}
\newcommand{\ri}{{\rm i}}
\newcommand{\rd}{{\rm d}}

\newcommand{\ba}{\begin{aligned}}
\newcommand{\ea}{\end{aligned}}
\newcommand{\ben}{\begin{eqnarray}\displaystyle}
\newcommand{\een}{\end{eqnarray}}

\def\tr{\hbox{tr}}

\def\tx{\tilde x}
\def\ty{\tilde y}
\newcommand{\ssq}[1]{{\sqrt{\sigma({#1})}}}

\newcommand{\figref}[1]{Fig.~\protect\ref{#1}}

\def\blfootnote{\xdef\@thefnmark{}\@footnotetext}
\long\def\symbolfootnote[#1]#2{\begingroup%
\def\thefootnote{\fnsymbol{footnote}}\footnote[#1]{#2}\endgroup}

\begin{document}

\begin{titlepage}

\begin{flushright}
BONN-TH-2007-07\\
CERN-PH-TH/2007-153\\
NEIP-07-03\\
\vspace{0.5cm}\end{flushright}

\centerline{\Large \bf Remodeling the B-model}

\medskip

\vspace*{3.0ex}

\centerline{\large \rm
Vincent Bouchard$^{a}$, Albrecht Klemm$^{b}$, Marcos Mari\~no$^{c}$ and Sara Pasquetti$^{d}$}

\vspace*{4.0ex}
\begin{center}
{\em $^a$ Jefferson Physical Laboratory, Harvard University\\
17 Oxford St., Cambridge, MA 02138}

\vspace*{1.8ex}
{\em $^b$Physikalisches Institut der Universit\"at Bonn, Nu\ss allee 12\\
D-53115 Bonn, Germany}

\vspace*{1.8ex}
{\em $^c$Department of Physics, CERN\\[.1cm]
Geneva 23, CH-1211 Switzerland}

\vspace*{1.8ex}
{\em $^d$ Institut de Physique, Universit\'e de Neuch\^atel\\
Rue A.  L. Breguet 1, CH-2000 Neuch\^atel, Switzerland
}

\symbolfootnote[0]{\tt bouchard@physics.harvard.edu,\
aklemm@physics.wisc.edu, \, marcos@mail.cern.ch, \,
sara.pasquetti@unine.ch}

\vskip 0.0cm
\end{center}

\centerline{\bf Abstract}
\medskip
We propose a complete, new formalism to compute unambiguously B-model open and closed amplitudes in local Calabi--Yau geometries, 
including the mirrors of toric manifolds. The formalism is based on the recursive solution of matrix models recently proposed by Eynard and Orantin. The resulting amplitudes are non-perturbative in both the closed and the open moduli. The formalism can then be used to study stringy phase transitions in the open/closed moduli space. At large radius, this formalism may be seen as a mirror formalism to the topological vertex, but it is also valid in other phases in the moduli space. We develop the formalism in general and provide an extensive number of checks, including a 
test at the orbifold point of $A_p$ fibrations, where the amplitudes compute the 't Hooft expansion of Wilson loops in lens spaces. We also use our formalism to predict the disk amplitude for the orbifold $\IC^3 / \IZ_3$.
\vskip 1cm

\noindent September 2007
\end{titlepage}

\tableofcontents
\baselineskip=15pt plus 3pt\parskip=9pt
\section{Introduction}

\subsection{Motivation}

Topological string theory is an important subsector of string theory with various physical and mathematical 
applications, which has been extensively investigated since it was first formulated. This has led to many different ways 
of computing topological string amplitudes, based very often on string dualities. Topological strings come in two types, the A-model and 
the B-model, which are related by mirror symmetry. The A-model provides a physical formulation of Gromov--Witten theory, while 
the B-model is deeply related to the theory of deformation of complex structures. Both models have an open sector whose boundary 
conditions are set by topological D-branes. 

The main advantage of the B-model is that its results are exact in the 
complex moduli ({\it i.e.} they include all $\alpha'$ corrections), which makes it possible to study various aspects of 
stringy geometry not easily accessible in the A-model. One can in
particular obtain results for the amplitudes far from the large radius limit, 
around non-geometric phases such as orbifold or conifold points. 

Sphere and disk amplitudes are given by holomophic integrals in the 
B-model geometry. In particular, sphere amplitudes are determined 
by period integrals over cycles; those were first calculated for the 
quintic Calabi--Yau threefold in \cite{cgpo}. For non-compact Calabi-Yau threefolds, the mirror geometry basically reduces to a Riemann
surface, and the disk amplitudes are given by chain integrals directly
related to the Abel-Jacobi map \cite{av,akv}. Note that disk amplitudes ending on 
the real quintic inside the quintic threefold have also been calculated \cite{walcher}, 
using a generalization of the Abel-Jacobi map~\cite{griffith}.

In contrast, B-model amplitudes $A_{h}^{(g)}$ at genus $g$ and with $h$ holes, on
worldsheets with $\chi<1$, have an anomalous, non-holomorphic dependence 
on the complex moduli which is captured by the holomorphic anomaly equations. 
These were first formulated in the closed sector in
\cite{bcov}, and  have been recently extended to the open sector in 
various circumstances \cite{emo,walchertwo,bt}. The holomorphic anomaly equations can be solved to 
determine the amplitudes, up to an {\it a priori} unknown  
holomorphic section over the moduli space --- the so called holomorphic
ambiguity ---  which puts severe restrictions on their effectiveness. 
Modular invariance of the amplitudes completely governs the non-holomorphic 
terms in the amplitudes~\cite{yy,hk,abk,hkq,emo} and reduces the problem of fixing the 
holomorphic ambiguity to a finite set of data for a given $g$ and $h$.    
Recently, boundary conditions have been found in the closed sector 
\cite{hk} (the so-called conifold gap condition and 
regularity at orbifold point) which fully fix these data
in many local geometries (like the Seiberg--Witten geometry or 
local $\IP^2$)~\cite{hk,gkmw}. In the compact case they allow 
to calculate closed string amplitudes to high, but finite genus (for example, $g=51$ for the 
quintic)~\cite{hkq}.   
 
For open string amplitudes the situation is worse: appropriate boundary 
conditions are not known, and the constraints coming from modularity 
are much weaker. In fact, it is not known how to supplement the holomorphic 
anomaly equations with sufficient conditions in order to fix the open string amplitudes.\footnote{This applies also to the case in which it has been argued 
that there is no open string moduli \cite{walcher}.}  

In view  of this, it is very important to have an approach to the B-model that goes beyond the framework of the holomorphic anomaly equations. 
In the local case (toric or not), such an approach was proposed in \cite{adkmv}, which interpreted the string field theory of the B-model (the Kodaira--Spencer theory) 
in terms of a chiral boson living on a ``quantum'' Riemann surface. However, the computational framework of \cite{adkmv} is only 
effective in very simple geometries, and in practice it is not easy to apply it even to backgrounds like local $\IP^2$. 

In \cite{mm2} it was argued that all closed and open topological string amplitudes on local geometries (including the 
mirrors of toric backgrounds) could be computed by adapting the recursive method of Eynard and Orantin \cite{eo} to 
the Calabi--Yau case. This method was obtained originally as a solution to the loop equations of matrix models, 
giving an explicit form for its open and closed amplitudes in terms of residue calculus on the spectral curve of the matrix model. 
The recursion solution obtained in this way can then be defined formally for {\it any} algebraic curve embedded in $\IC^2$. 
In \cite{mm2} it was argued that this more general construction attached to an arbitrary Riemann surface computes the amplitudes of the chiral 
boson theory described in \cite{adkmv}, and in particular that the formalism of \cite{eo} should give the solution to the 
B-model for mirrors of toric geometries, providing in this way 
an effective computational approach to the Kodaira--Spencer theory in the local case. 
Various nontrivial examples were tested in \cite{mm2} successfully. However, many important aspects of the B-model, like the 
phase structure of D-branes, as well as the framing phenomenon discovered in \cite{akv}, were not incorporated in the formalism of \cite{mm2}.

\subsection{Summary of the results}

In this paper we build on \cite{mm2} and develop a complete theory of the B-model for local Calabi-Yau geometries in the presence of 
toric D-branes. Our formalism is based on a modification of \cite{eo} appropriate for the toric case, and it leads to a framework where 
one can compute {\it recursively} and {\it univoquely} all the open and closed B-model amplitudes in closed form, non-perturbatively in the complex moduli, albeit 
perturbatively in the string coupling constant. In particular, our formalism incorporates in a natural way the more subtle aspects of D-branes (like 
framing) which were not available in \cite{mm2}. 

Moreover, the proposed formalism is valid at any point in the moduli space. Basically, once one knows the disk and the annulus amplitudes at a given point, one can generate recursively all the other open and closed amplitudes unambiguously. Thus this formalism goes beyond known approaches in open 
topological string theory, such as the topological vertex, as it allows 
to study closed and  open amplitudes not only in the large 
radius phase but also in non-geometric phases such as conifolds 
and orbifolds phases. 

\subsection{Outline}

In section 2 we review relevant features of mirror symmetry as well as open and closed topological string theory. We put a special emphasis on phase transitions in the open/closed moduli spaces, and on the determination of the corresponding open and closed mirror maps. Section 3 is the core of the paper, where we propose our formalism. We also explain how it can be used to compute amplitudes explicitly at various points in the moduli space. In sections 4 and 5 we get our hands dirty and do various checks of our formalism, for local geometries. In section 6 we study open and closed amplitudes in a non-geometric phase corresponding to the blow-down of local $\IP^1 \times \IP^1$. We check our results against expectation values of the framed unknot in Chern-Simons theory on lens spaces. We also propose a prediction for the disk amplitude of $\IC^3 / \IZ_3$, which corresponds to the orbifold phase in the moduli space of local $\IP^2$. Finally, we summarize and propose various avenues of research in section 7.

\subsection*{Acknowledgments}

It is a pleasure to thank Ron Donagi, Andy Neitzke, Tony Pantev, Cumrun Vafa and Johannes Walcher for helpful discussions. V.B., A.K. and S.P. would also like to thank the Theory Group 
at CERN for hospitality while part of this work was completed. The work of S.P. was partly supported by the Swiss National Science Foundation and by the European
Commission under contracts MRTN-CT-2004-005104. The work of V.B. was partly supported by an NSERC postdoctoral fellowship.

\section{Toric Calabi-Yau threefolds with branes}

In this section we introduce basic concepts of mirror symmetry 
and topological string theory for non-compact toric Calabi-Yau threefolds
with Harvey-Lawson type special Lagrangians. In particular, we discuss the 
target space geometry of the A- and the B-model as well as 
their moduli spaces. We also examine period integrals on the B-model side, which give 
the flat coordinates as well as the closed genus zero
and disk amplitudes. 

\subsection{Mirror symmetry and topological strings on toric Calabi-Yau threefolds}

\subsubsection{A-model geometry}  
\label{amodel} 

We consider the A-model topological string on a (non-compact) toric Calabi-Yau threefold, which can be 
described as a symplectic quotient $M=\mathbb{C}^{k+3}//G$, 
where $G=U(1)^k$~\cite{cox}. Alternatively, $M$ may be viewed physically as 
the vacuum field configuration for the complex scalars $X_i$, $i=1,\ldots,k+3$
of chiral superfields in a 2d gauged linear, $(2,2)$ supersymmetric 
$\sigma$-model, transforming as $X_i\rightarrow e^{ i Q_i^\alpha  \epsilon_\alpha} X_i$, 
$Q_i^\alpha\in \mathbb{Z}$, $\alpha=1,\ldots, 
k$ under the gauge group $U(1)^k$~\cite{wittenphases}. 
Without superpotential, $M$ is determined by the $D$-term constraints 
\begin{equation} 
D^\alpha=\sum_{i=1}^{k+3} Q_i^\alpha|X_i|^2=r^\alpha, \quad \alpha=1,\ldots, k 
\label{dterm} 
\end{equation}
modulo the action of $G=U(1)^k$. The $r^\alpha$ are the K\"ahler parameters and
$r^\alpha\in \mathbb{R}_+$ defines a region in the K\"ahler cone. For this to be 
true $Q_i^\alpha$ have to fullfill additional constraints and for $M$ 
to be smooth, field configurations for which the dimensionality 
of the gauge orbits drop have to be excluded.

The Calabi-Yau condition 
$c_1(TX)=0$ holds if and only if the chiral $U(1)$ anomaly is cancelled, that is~\cite{wittenphases}
\begin{equation}  
\sum_{i=1}^{k+3} Q_i^\alpha=0, \qquad \alpha = 1, \ldots, k .
\label{anomaly}
\end{equation}  
Note from \eqref{dterm} that negative $Q_i$ lead to non-compact directions in $M$, 
so that all toric Calabi-Yau manifolds are necessarily non-compact.

View the $\mathbb{C}$'s with coordinates $X_k=|X_k|\exp(i\theta_k)$ as $S^1$-fibrations over $\mathbb{R}_+$. Then $M$ can be naturally viewed as 
a $T^3$-fibration over a non-compact convex and linearly bounded subspace $B$ in $\mathbb{R}^3$ 
specified by \eqref{dterm}, where the $T^3$ is parameterized by the three directions
in the $\theta$-space
. The condition 
\eqref{anomaly} allows an even simpler picture, capturing the geometry 
of $M$ as a $\mathbb{R}_+\times T^2$ fibration over $\mathbb{R}^3$. In this picture, the toric threefold $M$ is constructed by gluing together $\IC^3$ patches. In 
each patch, with coordinates $(z_1,z_2,z_3)$, we can define, instead of $r_{\alpha_i}=|z_i|^2$ ---
which would lead to the above picture --- the three following hamiltonians    
\begin{equation}
r_{\alpha}=|z_1|^2-|z_2|^2,\qquad r_{\beta}=|z_3|^2-|z_1|^2,\qquad r_{{\rm R}}=
{\rm Im}(z_1z_2z_3)\ . 
\label{hamiltonian}
\end{equation}
The $r_{l}$ parameterize the base $\mathbb{R}^3$ and generate flows 
$\delta_{r_l} x_k=\{r_l,x_k\}_\omega$, whose orbits define the fiber. 
It is easy to see that $r_{\alpha},r_{\beta}$ generate $S^1$'s and
$r_{{\rm R}}$, which is only well defined due to \eqref{anomaly}, generates
$\mathbb{R}_+$.      

The toric graph $\Gamma_M$ describes the degeneration locus of the 
$S^1$ fibers. In $B$, $|X_i|\ge 0$, therefore $B$ is bounded by $|X_i|=0$. 
The latter equations define two-planes in $\mathbb{R}^3$ whose normal vectors obey 
\be
\sum_{i=1}^{3 + k} Q^\alpha \vec n_i=0.
\ee
Clearly, the $S^1$ parameterized 
by $\theta_i$ vanishes at $|X_i|=0$; and over the line segments 
\be
L_{ij}=\{|X_i|=0\}\cap \{|X_j|=0\},
\ee
two $S^1$'s shrink to zero. 
If $L_{ij}$ is a closed line segment in $\partial B$ the open $S^1$ 
bundle over it make it a $\mathbb{P}^1\in M$, while 
if $L_{ij}$ is half open in  $\partial B$ it represents a 
non-compact line bundle direction $\mathbb{C}$. 

So far we have defined the planes $|X_i|=0$ only up 
to parallel translation. Their relative location is determined by 
the K\"ahler parameters, simply by the condition that the length
of the closed line segments $L_{ij}$ is the area of the corresponding 
$\mathbb{P}^1$. Condition \eqref{anomaly} and the $T^2$  
fibration described obove makes it possible to project 
all $L_{ij}$ into $\mathbb{R}^2$ without losing information about the 
geometry of $M$. This is how one constructs the two-dimensional toric graph 
$\Gamma_M$ associated to $M$.

\subsubsection{B-model mirror geometry}
\label{bmodel}

The mirror geometry $W$ to the above non-compact toric Calabi-Yau threefold $M$ was
constructed by~\cite{Hori:2000kt}, extending~\cite{Katz:1996fh, Bat}.

Let $w^+, w^- \in \mathbb{C}$. We further define homogeneous coordinates 
$x_i=:e^{y_i} \in \IC^*$, $i=1,\ldots, k+3$ with the property 
$|x_i|=\exp(-|X_i|^2)$; they are identified under the $\mathbb{C}^*$-scaling $x_i\sim \lambda x_i$, $i=1,\ldots,k+3$, 
$\lambda\in \mathbb{C}^*$. The mirror geometry $W$ is then given by 
\begin{equation} 
w^+ w^- =\sum_{i=1}^{k+3} x_i \ ,
\label{mirrorgeometry} 
\end{equation}
subject to the exponentiated $D$-terms contraints, which  
become 
\begin{equation} 
\prod_{i=1}^{k+3} x_i^{Q_i^\alpha}=e^{-t^\alpha}=q^\alpha,\quad \alpha=1,\ldots, k \ . 
\label{exponenteddterm}
\end{equation} 
Note that these relations are compatible with the $\lambda$-scaling because of the Calabi-Yau condition. The 
parameters $t^\alpha=r^\alpha+i\theta^\alpha$ are the complexifications
of the K\"ahler parameters $r^\alpha$, using the $\theta^\alpha$-angles of the $U(1)^k$ 
group. 

After taking the $\lambda$-scaling and \eqref{exponenteddterm} into account the 
right-hand side of the defining equation \eqref{mirrorgeometry} can be parameterized by two variables 
$x=\exp(u),y=\exp(v) \in \IC^*$. In these coordinates the mirror geometry $W$ becomes
\begin{equation} 
\label{mirrorcurve}
w^+ w^-=H(x,y; t^\alpha),
\end{equation}
which is a conic bundle over $\mathbb{C}^* \times \IC^*$, where the conic fiber 
degenerates to two lines over the (family of) Riemann surfaces $\S: \{H(x,y; t^\alpha)=0 \} \subset \IC^* \times \IC^*$.\footnote{Note that for brevity in the following we will always talk about the Riemann surface $\S$; it will always be understood that $\S$ is in fact a family of Riemann surfaces parameterized by the K\"ahler parameters $t^\alpha$.} The holomorphic volume form on $W$ is given by
\begin{equation}
\Omega={d w d x  d y\over w x y}.
\end{equation}

As an algebraic curve embedded in $\IC^* \times \IC^*$, the Riemann surface $\S$ has punctures, hence is non-compact. The fact that it is embedded in $\IC^* \times \IC^*$ rather than $\IC^2$ like the usual
specialization  of a compact Riemann surface embedded in projective space  to an affine coordinate patch will be crucial for us. Note that the Riemann surface $\S$ is most easily visualized by fattening the toric diagram $\Gamma_M$ associated to the mirror manifold $M$; the genus $g$ of $\S$ corresponds to the number of closed meshes in $\Gamma_M$, and the number of punctures $n$ is given by the number of 
semi-infinite lines in $\Gamma_M$. It is standard to call the Riemann surface $\S$ embedded in $\IC^* \times \IC^*$ the \emph{mirror curve}.          

It is important to note that the reparameterization group $G_\S$ of the mirror curve $\S$ is 
\be
G_\S = SL(2,\IZ) \times \begin{pmatrix} 0 & 1 \\ 1 & 0 \end{pmatrix},
\ee
which is the group of $2 \times 2$ integer matrices with determinant $\pm 1$. This is the group that preserves the symplectic form 
\be 
\left | {d x \over x} \wedge {d y \over y}\right|
\ee 
on $\IC^* \times \IC^*$. The action of $G_\S$ is given by
\begin{equation} 
(x,y) \mapsto (x^a y^b, x^c y^d), \qquad \left(\begin{array}{cc} a& b\\ c& d\end{array}\right)\in G_\S.
\label{sl2z}
\end{equation}

\subsubsection{Open string mirror symmetry}

\label{openms}

In this work we are interested in closed and open topological string 
amplitudes, hence we must consider branes, which are described in the 
A-model by special Lagrangian submanifolds. The Lagrangian submanifolds 
that we will be interested in were constructed by~\cite{av}, as a generalization of 
Harvey-Lawson special Lagrangians~\cite{HarveyLawson} 
in $\mathbb{C}^3$. 

Consider a toric Calabi-Yau threefold $M$ constructed as a symplectic quotient as above, and denote by 
\begin{equation} 
\omega=\frac{1}{2}\sum_{k=1}^3 d |X_k|^2\wedge d \theta_k 
\end{equation} 
the canonical symplectic form. The idea is to determine a 
non-compact subspace $L \subset M$ of three real dimensions by specifying a linear subspace $V$ in the base
\begin{equation}
\sum_{i=1}^{k+3} q_i^\alpha |X_i|^2=c^\alpha, \quad \alpha=1,\ldots,r     
\end{equation}
and restricting the $\theta_k$ so that $\omega|_L=0$.  One shows that 
$L$ becomes special Lagrangian with respect to $\Omega=d z_1\wedge d z_2\wedge d z_3$ 
in each patch if and only if 
\be
\sum_{i=1}^{k+3} q_i^\alpha=0,\qquad \alpha=1,\ldots,r.
\ee    

The relevant case for us is $r=2$, i.e. $V=\mathbb{R}_+$ and $L$ is an 
$S^1\times S^1$-bundle over it. In a given patch,  
the restriction  $\omega|_L=0$ means that 
\be
\sum_{i=1}^3 \theta_i(z_i)=0\ { \rm mod}\ \pi .
\ee
For one value of the $\theta$-sum the Lagrangian is generically not 
smooth at the origin of $\IR_+$. It can be made smooth by ``doubling,'' which 
is done by allowing for instance $\sum_{i=1}^3 \theta_i(z_i)=0$ and 
$\sum_{i=1}^3 \theta_i(z_i)=\pi$~\cite{HarveyLawson}. 
If $V$ passes through the locus in the base where one $S^1$ shrinks to zero,  
$L$ splits into $L_\pm$, each of which have topology $\mathbb{C}\times S^1$, 
where $\mathbb{C}$ is a fibration of the vanishing $S^1$ over $\mathbb{R}_+$.  
$L_+$ (or $L_-$) is the relevant special Lagrangian. To make the 
notation simpler we denote $L_+$ by $L$ henceforth. It has 
$b_1(L)=1$ and its complex  open modulus is given by the 
size of the $S^1$ and the Wilson line of the 
$U(1)$ gauge field around it. Pictorially, it can be described as 
``ending on a leg of the toric diagram $\Gamma_M$ of $M$,'' since the half open line $l$ defining
$L$ must end on a line $L_{ij}$ in $\Gamma_M$. We refer the reader to the figures in sections 4 and 5 for examples of toric diagrams with branes.

Under mirror symmetry, the brane $L$ introduced above maps to a one complex dimensional holomorphic submanifold of $W$, given by
\begin{equation}
\label{bbrane}
H(x,y) = 0 = w^-.
\end{equation}
That is, it is parameterized by $w^+$, and its moduli space corresponds to the mirror curve $\S \subset \IC^* \times \IC^*$ ($w^+=0$ 
corresponds to the equivalent brane $L_-$). 

\subsubsection{Topological open string amplitudes}

Let us now spend a few words on topological open string theory to clarify the 
objects that we will consider in this paper.

In the A-model, topological open string amplitudes can be defined by counting (in an appropriate way) the number of holomorphic maps 
from a Riemann surface $\Sigma_{g,h}$ of genus $g$ with $h$ holes, to the Calabi--Yau target, satisfying the condition that the boundaries map 
to the brane $L$. Assuming for simplicity that $b_1(L)=1$, the topological class of these maps is labeled by genus $g$, 
the bulk class $\beta \in H_2(X, L)$ and the winding numbers $w_i$, 
$i=1, \cdots, h$, specifying how many times the $i$-th boundary wraps around the one-cycle in $L$. We can thus form the generating functionals
\be
F_{g,w}(Q)=\sum_{\beta\in H_2(X,L)} N_{g,w,\beta} \, \re^{-\beta\cdot t}, 
\ee
where $N_{g,w,\beta}$ are open Gromov--Witten invariants counting the maps in the topological class labeled by $g$, $w=(w_1, \cdots, w_h)$, and $\beta$. 
It is also convenient to group together the different boundary sectors with fixed $g,h$ into a single generating functional $A_h^{(g)} (z_1, \cdots, z_h)$ defined as
\be
\label{opena}
A^{(g)}_h(z_1, \cdots, z_h)=\sum_{w_i \in \IZ} F_{g,w}(Q) z_1^{w_1} \cdots z_h^{w_h}. 
\ee
Here, the variables $z_i$ are not only formal variables. From the point of view of the underlying physical theory, they are 
\emph{open string parameters} which parameterize the moduli space of the brane.

In the B-model, as discussed earlier the moduli space of the brane is given by the mirror curve $\Sigma$ itself. The open string parameter $z$ hence corresponds to a variable on the mirror curve $\Sigma$ (take for example the variable $x$). That is, fixing what we mean by open string parameter corresponds to fixing a parameterization of the embedding of the Riemann surface $\Sigma$ in $\IC^* \times \IC^*$; in other words, it corresponds to fixing a projection map $\S \to \IC^*$ (the projection onto the $x$-axis in our case). Different parameterizations will lead to different amplitudes. 
Once the open string parameter $x$ is fixed, the disk amplitude is simply given by
\be
A_1^{(0)} = \int \log y {\rd x \over x},
\label{diskamp}
\ee
as will be explained in more details in the following sections.

To fully understand open topological strings we need to include the notion of \emph{framing} of the branes. 
The possibility of framing was first discovered in the context of A-model open string amplitudes 
in \cite{akv}. It is an integer choice $f\in \mathbb{Z}$ associated to a brane, which has to be made 
in order to define the open amplitudes.

Framing has various interpretations. In the A-model, it corresponds to an
integral choice of the circle action with respect to which the localization 
calculation is performed~\cite{kl}. It can also be understood from the point of view of large $N$ duality.
A key idea in the large $N$ approach is to relate open (and closed) 
string amplitudes  to knot or link invariants in the Chern-Simons 
theory on a special Lagrangian cycle. As is well known the calculation of 
the Chern-Simons correlation functions requires a choice of the normal bundle 
of each knot. The framing freedom lies 
in a twist of this bundle, again specified by an integer  $f\in \mathbb{Z}$.

We also want to understand framing from the B-model point of view. Recall that the moduli space of the brane is given by the mirror curve $\Sigma$. As explained above, fixing the location on the brane on the A-model corresponds to fixing a parameterization of $\Sigma$. It turns out that there is a one-parameter subgroup of the reparameterization group $G_\S$ of $\Sigma$ which leaves the location of the brane invariant; these transformations, which depend on an integer $f \in \mathbb{Z}$, correspond precisely to the B-model description of framing \cite{akv}. More precisely, these transformations, which we will call \emph{framing transformations}, are given by
\be
(x,y) \mapsto (x y^f, y), \qquad f \in \IZ.
\label{framtrans}
\ee
As a result, fixing the location and the framing of the brane on the A-model side corresponds to fixing the parameterization of the mirror curve on the B-model side.


\subsection{Moduli spaces, periods and flat coordinates}
\label{modulispace}
In this section we discuss the global picture of the
open/closed moduli space of the A- and the B-model. 
We introduce the periods, which give us the open and closed 
flat coordinates, as well as the disk amplitude and the closed genus zero 
amplitude.
 
\subsubsection{Moving in the moduli space}
\label{moving}

Mirror symmetry identifies the stringy K\"ahler moduli space of $M$ with the complex structure moduli space of $W$, which are the A- and B-model closed string moduli spaces. Recall that generically, the stringy K\"ahler moduli space of $M$ contains various phases corresponding to topologically distinct manifolds. Hence moving in the A-model closed string moduli space implies various topologically-changing phase transitions corresponding to flops and blowups of the target space. In fact, since we are interested in open topological strings, we want to consider the open/closed string moduli space, which also includes the moduli space associated to the brane.

The B-model provides a natural setting for studying transitions in the open/closed string moduli space. Usually in mirror symmetry, we identify the A- and B-model moduli spaces locally by providing a mirror map, for example near large radius and for outer branes. However, in the following we will propose a B-model formalism to compute open/closed amplitudes which can be applied \emph{anywhere} in the open/closed string moduli space. Hence, to unleash the analytic power of this new B-model description one wishes to extend the identification between the moduli space to cover all regions of the open/closed string moduli space.

In the B-model one simply wants to cover a suitable compatification of the open/closed string moduli 
space with patches in which we can define convergent expansions of the topological string amplitudes in local 
\emph{flat coordinates}. The latter are given by a choice of A-periods integrals, while the dual B-periods 
can be thought of as conjugated momenta. The closed string flat coordinates are given 
by integrals over closed cycles, while the open string flat coordinates are integrals 
over chains. 

Let us first discuss the closed string flat coordinates. If the genus of the B-model mirror curve is greater than one, one has non-trivial monodromy of 
the closed string periods. By the theory of solution to  differential equations with regular singular loci (normal 
crossing divisors), which applies in particular 
to periods integrals, the closed string moduli space can 
be covered by hyper-cylinders around the divisors with 
monodromy. The local holomorphic expansion of 
the amplitudes has to be invariant under the local 
monodromy around the corresponding divisor. 
In particular, this requires different choices 
of flat coordinates, or A-periods, in different regions in moduli space. These different choices of periods are related
by symplectic ${\rm Sp}(2g,\mathbb{C})$ transformations, 
\emph{i.e.} by changes of polarization.
Invariance of the physical topological string amplitudes under the full monodromy group 
requires a non-holomorphic extension of the amplitudes and 
forces the closed string parameters to 
appear in terms of modular forms. 

In contrast there 
is no monodromy action on the open string flat coordinates. 
As a consequence, the amplitudes are in general rational functions of the open string parameters, and no non-holomorphic extension is needed to make the results modular. That is, there is no holomorphic anomaly equation involving the complex conjugate of an open string modulus. The situation for the open string moduli is hence similar 
to the closed string moduli for genus $0$ mirror curves (for example the mirror of the resolved conifold $M = {\cal O}(-1)\oplus {\cal O}(-1)\rightarrow 
\mathbb{P}^1$), where there is no non-trivial monodromy. In such cases the holomorphic anomaly equations for the closed string moduli can be trivialized and the amplitudes are rational functions of the moduli.  

Let us now discuss the main features of the phase
transitions\footnote{Note that the term ``phase transitions" is 
inspired from the classical A-model. In the B-model, the correlation 
functions are smoothly differentiable except at complex dimension 
one loci, so there are strictly speaking no phase boundaries.} between patches in the open/closed string moduli space in order of their 
complexity.  In the easiest case adjacent patches 
are related by transitions merely in the open string moduli 
space. In the A-model these are referred to as open string phase 
transitions and correspond to moving the base of the special 
lagrangian submanifold over a vertex in the toric diagram, for 
example  from an outer to an inner brane, see below. 
In the B-model they correspond simply to reparameterizations 
of the mirror curve $\S$ by an element in $G_\S$. More precisely, this type of phase transition is described by the reparameterization
\be
(x,y) \mapsto \left( {1 \over x},  {y \over x} \right)
\label{phasetrans}
\ee 
of the mirror curve $\Sigma$ --- we will explain this in the next section. Since the amplitudes are rational functions in the open 
string moduli, there is no non-trival analytic 
continutation required and the amplitudes can be readily 
transformed. 

The next type of transitions consists in closed string transitions between different large radius
regions. In the A-model on non-compact toric Calabi-Yau threefolds those are all related to flops of
$\mathbb{P}^1$ (in our examples they occcur only in the Hirzebruch surface 
$F_1$). In these cases, the new flat closed string coordinates
are given linearly in terms of the old ones and in particular 
the symplectic transformation in ${\rm Sp}(2g,\mathbb{Z})$ 
is trivial, in the sense that it does not exchange 
the A- and B-periods. The closed string parameters can 
be fixed in each large radius patch by the methods of 
\cite{akv}, which are reviewed in the next section. The rather 
mild changes in the amplitudes can be described 
by wall crossing formulas.
   
The more demanding transitions are the ones 
between patches which require a non-trival 
${\rm Sp}(2g,\mathbb{C})$ transformation of the periods. 
The typical example is the expansion near 
a conifold point. Here a B-cycle --- in the choice of periods
at large radius --- becomes small and will serve as a flat 
coordinate near the conifold point, while a cycle corresponding to a 
flat coordinate at infinity aquires a logarithmic 
term and will serve as dual momenta. In the A-model 
picture we enter a non-geometric phase, in which the 
$\alpha'$-expansion of the $\sigma$-model breaks 
down. In the B-model we are faced with  
the problem of analytic continuation and change of 
polarization when we transform the amplitudes, which involve modular transformations.

Another interesting patch is the one of an orbifold 
divisor, i.e. one with a finite monodromy around it. 
This is likewise a region where the original geometric 
description breaks down due to a vanishing volume. 
However here we have a singular geometric description by a geometric 
orbifold. For example, for the 
${\cal O}( -3)\rightarrow \mathbb{P}^2$ geometry, in the 
limit where the $\IP^2$ shrinks to zero size we get simply the $\mathbb{C}^3/\mathbb{Z}_3$ 
orbifold. Enumerative A-model techniques (orbifold Gromov-Witten invariants) have been 
developed to calculate closed string invariants on orbifolds, and in these phases we can still compare the closed B-model results with Gromov-Witten calculations on the A-model side. The behavior of the closed string amplitudes under this type of transition has been studied in \cite{abk}. In this paper, we will start investigating open amplitudes on orbifolds, which do not have, as far as we know, a Gromov-Witten interpretation. In particular, 
we will calculate the disk amplitude for $\IC^3 / \IZ_3$ in section \ref{c3z3orbifold}.

Let us now describe in more detail the first type of phase transitions, involving only open string moduli.


\subsubsection{Open string phase structure}
\label{openstringphases}
Here we introduce classical open string coordinates and discuss 
the phases of the open string moduli, which 
arise when we ``move'' the Lagrangian submanifold over a vertex in the toric 
diagram. 

First, note that the open string variables generically get corrected by closed string instanton effects, when the latter are present and have finite volume; we will study this in the next section. However, the open string phase 
structure can already be understood directly in the large volume 
limit where the instanton corrections are suppressed. Hence, we will not bother for now with the instanton corrections; our analysis carries over readily 
to the instanton corrected variables. 

Recall from section \ref{amodel} that closed line segments in the toric diagram $\Gamma_M$ correspond to compact curves, while half-open lines correspond to non-compact curves. Now, as explained in section \ref{openms}, the half open line $l$ defining the Lagrangian submanifold $L$ must end on a line $L_{ij}$ in $\Gamma_M$. Phase transitions in the open string moduli space then occur between Lagrangian submanifolds ending on half-open lines and Lagrangian submanifolds ending on closed line segments. One refers to the 
former as {\it outer branes} and to the latter as {\it inner branes}. 

\begin{center}
\begin{figure}[thbp]
\begin{center}
\includegraphics[width=6in]{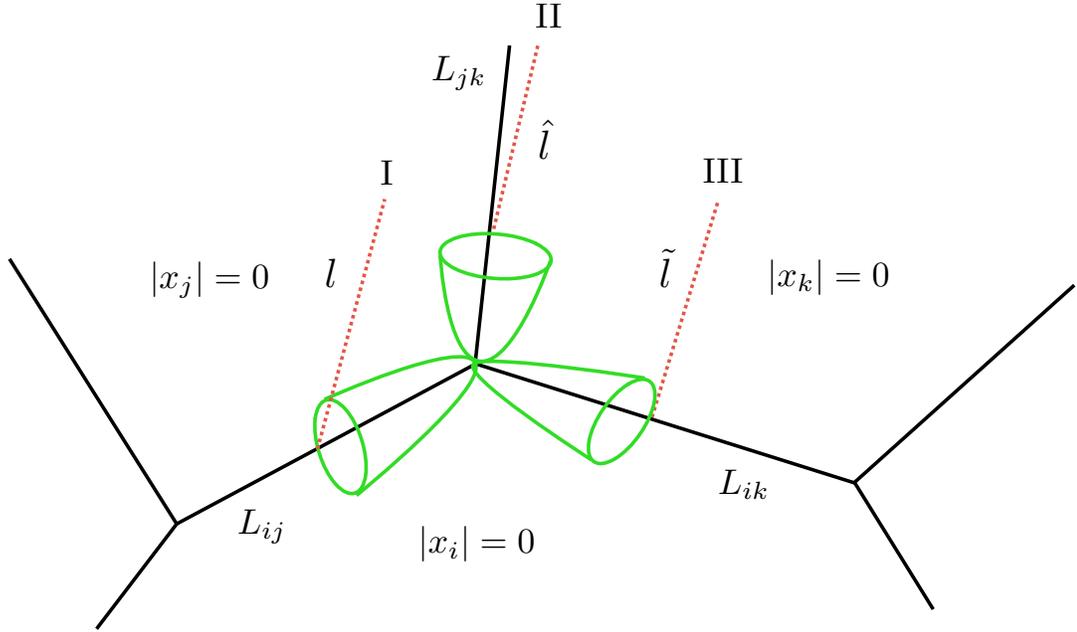}
\end{center}
\caption{Open string phase structure.}
\label{fig:1}
\end{figure}
\end{center}

Only maps which are equivariant 
with respect to the torus action contribute to the open string amplitudes. This means that disks 
must end on a vertex at one end of the line $L_{ij}$ intersecting $l$. Let this vertex be the locus 
where $|X_i|=|X_j|=|X_k|=0$. Branes ending on the three lines 
$L_{ij}$, $L_{ik}$ and $L_{jk}$ meeting up at this vertex correspond to three different phases $I$, $II$ and $III$  
in the open string moduli space. The geometry of the open string phase structure is shown in figure \ref{fig:1}.

In phase I we can describe  $l$ by the equations
\begin{equation}
\begin{array}{rl} 
|X_j|^2-|X_i|^2&=0,\\
|X_k|^2-|X_i|^2&=c_{r} , \qquad r > c_r> 0,
\end{array}
\label{braneI} 
\end{equation}
where $r$ is the K\"ahler parameter of the $\mathbb{P}^1$ related to $L_{ij}$ 
and  
\be
c_r=\int_{S^1} H,
\ee
where $\rd H=\omega$ parameterizes the size of the disk $D$, hence the radius 
of the $S^1=\partial D$. 

Recall that on the B-model side, the choice of location (or phase) of the brane corresponds to a choice of parameterization of the mirror curve $\Sigma$ defined by $H(x,y) = 0$. Generically, we can find the good parameterization of the curve as follows. We first use the fact from mirror symmetry (see section \ref{bmodel}) that by definition,
\be
|x_i| = {\rm exp} \left( - | X_i |^2 \right),
\ee
to rewrite the equations \eqref{braneI} fixing the location of the brane in terms of the $\IC^*$-variables $x_i$. We then use the $\IC^*$-rescaling to fix one of them to $1$, and we choose $y$ to be the $\IC^*$-variable which goes to $1$ on the brane, and $x$ to be the variable parameterizing the location of the brane on the edge (\emph{i.e} $|x| = \re^{c_r}$). $x$ becomes the open string parameter introduced earlier. Note that there is an ambiguity in this choice of parameterization; since $y=1$ on the brane, we can reparameterize the variable $x \mapsto x y^f$ for any integer $f \in \IZ$ without changing the discussion above. But since we change the meaning of the open string parameter $x$, we in fact change the physical setup and the open amplitudes. This ambiguity precisely corresponds to the framing of the brane, and the transformation $x \mapsto x y^f$ is the framing transformation introduced in \eqref{framtrans}.

For example, in phase I, the good choice of parameterization corresponds to first scaling $x_i = 1$, and then identifying $y:= x_j = x_j/x_i$ and $x := x_k = x_k/x_i$. Indeed, the first equation in \eqref{braneI} says that $y=1$ on the brane, 
while the second equation identifies 
\be
x=\exp\left(c_r+i \int_{S^1} A \right),
\ee
where we complexified the disk size $c_r$ by the Wilson line. $x$ hence agrees in the 
large K\"ahler parameter limit with the open string parameter, which appears in the 
superpotential. In fact, the superpotential --- or disk amplitude --- is given by the Abel-Jacobi map on $H(x,y)=0$, as a curve embedded in $\IC^* \times \IC^*$, with respect to the restriction of the holomorphic volume form $\Omega$ to the mirror curve: 
\begin{equation} 
A_1^{(0)}(x)=\int_{x^*}^x \log y {\rd x \over x} , 
\label{superpotential}
\end{equation} 
\emph{i.e.} $x \partial_x A_1^{(0)}= \log y(x)$, with $y(x)$ a suitable branch of the solution of $H(x,y)=0$. This gives the formula for the disk amplitudes presented earlier in \eqref{diskamp}.
Note that we could also parametrize $l$ in this phase by 
\begin{equation}
\begin{array}{rl} 
|X_i|^2-|X_j|^2&=0,\\
|X_k|^2-|X_j|^2&=c_{r} , \qquad r > c_r> 0, 
\end{array}
\label{brane} 
\end{equation}
which leads to parameters $x' = x y^{-1}$ and $y'= y^{-1}$.

In phase II the brane $\hat l$ can be descibed by the equation 
\begin{equation}
\begin{array}{rl} 
|X_j|^2-|X_k|^2&=0,  \\
|X_i|^2-|X_k|^2& =c_{\hat r},\qquad  c_{\hat r}> 0\ .  
\end{array}
\label{brane} 
\end{equation}
We fix the parameterization of the mirror curve by $x_k=1$, $\hat y:= x_j = x_j/x_k$ and $\hat x:= x_i = x_i/x_k$, so that the open string parameter is $\hat x$ and the superpotential is \eqref{superpotential} with hatted variables. The relation to the previous parameters in phase I is $\hat x = x^{-1}$ and $\hat y = y x^{-1}$; this is the origin of the phase transformation proposed in \eqref{phasetrans}. 
Again, we can also parametrize $\hat l$ by 
\begin{equation}
\begin{array}{rl} 
|X_k|^2-|X_j|^2&=0,  \\
|X_i|^2-|X_j|^2&=c_{\hat r},\qquad  c_{\hat r}> 0, \ 
\end{array}
\label{brane} 
\end{equation} 
and get $\hat y'= x y^{-1}$ and  $\hat x'= y^{-1}$.  
      
Similarly, in phase III we can parameterize $\tilde l$ in two different ways, and introduce variables $\tilde y = x^{-1}$ and $\tilde x = x^{-1} y$, or $\tilde y' = x$ and $\tilde x' = y$. In this phase, $\tilde r$ is the K\"ahler parameter 
of the $\mathbb{P}^1$ related to $L_{ik}$. Note however that different $L_{nm}$ 
can describe  $\mathbb{P}^1$'s in the same K\"ahler class. Standard toric techniques
allow to read the equivalences from the charge vectors $Q^\alpha$.   

\subsection{The open and closed mirror maps}
\label{mirrormap}
 
We discussed in the previous section the phase structure of the open/closed moduli space. Here we discuss in detail how to find the flat coordinates (or open and closed mirror maps) in various phases in the moduli space.  As an example
we consider ${\cal O}(-3) \rightarrow \mathbb{P}^2$, which is the 
simplest non-compact toric Calabi-Yau with non-trivial monodromy on the closed 
string moduli. 
 
\subsubsection{Closed flat coordinates}
The closed string mirror map is given by finding {\it flat coordinates} $T^\alpha$, 
$\alpha=1,\ldots,k$ on the complex structure moduli space, which are mapped to the complexified K\"ahler parameters. The flat coordinates are generically defined by 
\be
T^\alpha={X^\alpha \over X^0},
\ee
where the $X^\alpha$ are the A-periods
\be
\label{Aclosedperiods}
X^\alpha = \int_{A^\alpha} \Omega
\ee
of the holomorphic volume form $\Omega$, and $(A^\alpha, B_\alpha)$ 
is a symplectic basis of three-cycles.
Special geometry guarantees the existence of a holomorphic function 
$F(X^\alpha)$ of degree $2$ --- the so-called prepotential --- such that the B-periods 
are
\be
F_\alpha = \frac{\partial F}{\partial X^\alpha} = \int_{B_\alpha} \Omega\ .
\label{Bclosedperiods}
\ee
 
Fixing the flat coordinates involves a choice of basis $(A^\alpha, B_\alpha)$; it is well known that the choice of A-periods (and the B-periods, i.e a polarization) is 
uniquely fixed at the point(s) of maximal unipotent monodromy 
${\underline q}=0$, which are mirror dual to the large radius 
points in the stringy K\"ahler moduli space. This fixes the 
closed mirror map at these large radius points.

To be more precise, in the paramerization of the complex moduli $q^\alpha=e^{-t^\alpha}$ 
determined by the Mori cone --- spanned by the charge vectors $Q^\alpha$ --- these periods 
are singled out by their leading behaviour: 
\be
X^0=1+{\cal O}(q), \qquad X^\alpha(q)=\log(q^\alpha)+ {\cal O}(q).
\ee 
In the non-compact cases there is a further simplification. First, $X^0 = 1$, and 
\begin{equation}
T^\alpha=X^\alpha=\frac{1}{2 \pi i}\int_{A^\alpha}\lambda .
\label{closedmirrormap}
\end{equation}
The period $F_0$ is absent and the dual periods are given by
\begin{equation}
F_\alpha= \frac{\partial F}{\partial T^\alpha}=\frac{1}{2 \pi i}\int_{B^\alpha}\lambda ,
\label{dualclosedperiods}
\end{equation}
where $(A^\alpha, B_\alpha)$ is now a canonical basis of one-cycles on the mirror curve $\Sigma$, and $\lambda$ is the meromorphic one form
\be
\lambda = \log y {\rd x \over x}
\ee 
on $\Sigma$, which is the local limit of $\Omega$.  

In the $A$-model picture, the flat coordinate $T^\alpha$ is the mass associated with 
a $D2$ brane wrapping the curve ${\cal C}_\alpha\in H_2(M,\mathbb{Z})$.  
At a large radius point, it is  given by the complexified volume 
\be
t^\alpha=\int_{{\cal C}_\alpha} \omega+i B.
\ee
However, it is well known that it receives closed string worldsheet 
instanton corrections if the size of ${\cal C}_\alpha$ is of the order 
of the string scale; the corrected volume
\be
T^\alpha=t^\alpha+{\cal O}(e^{-t^\alpha}),
\ee
is the flat coordinate, which reduces in the local case to (\ref{closedmirrormap}).
 
\subsubsection{Open flat coordinates}

The open string modulus is given by $x = \re^u$, which is a variable on the mirror curve $\Sigma$ defined by the equation $H(x,y)=0$. In this section we will sometimes use the variables $u$ and $v$ instead of the $\IC^*$-variable $x$ and $y$, which are defined by the exponentiation $x= \re^u$, $y = \re^v$. Hopefully no confusion should occur.

It was argued in \cite{akv} that in the A-model, the open string modulus $u$ measures the tension 
\be
\Delta W=W(x_3=-\infty)-W(x_3=\infty)
\ee
of a domain-wall made from a $D4$-brane wrapping the disk of classical size $u$ 
and extending at a point on the $x_3$-axis, say $x_3=0$, over the subspace $M_{2,1}$ of 
the four-dimensional Minkowski space $M_{3,1}$. In the large radius limit, this can be 
identified on the $B$-model side with the integral 
\be
\frac{1}{2 \pi i} \int_ {\alpha_u} v(u)  \rd u = \frac{1}{2 \pi i} \int_ {\alpha_u} 
y(x){\rd x \over x},
\ee 
where ${\alpha_u}$ 
is a not a closed cycle but rather a chain over which $v$ 
jumps by $2 \pi i$.  In analogy with (\ref{closedmirrormap}),
one expects that 
\begin{equation}
U=\frac{1}{2 \pi i}\int_{\alpha_{u} } \lambda
\label{openmirrormap}
\end{equation}
is the exact formula for the flat open string parameter $U$, which includes 
all instanton corrections.

Note that the above indeed depends on a choice of parameterization 
of the curve, which defines the location/phase and framing of the brane. 
In principle, the chain  ${\alpha_u}$ and the integral (\ref{openmirrormap}) 
can be obtained for branes in any phases. However, in practice, it turns 
out to be easier to start with outer branes, and use the open moduli phase 
transitions explained in the previous section, which relate the coordinates 
in various phases, to extract the flat open string parameters in other phases. 
Finally, note also that it is straightforward to show that both 
(\ref{closedmirrormap}) and (\ref{openmirrormap}) receive only 
closed string worldsheet instanton corrections.

The open string disk amplitude $A^{(0)}_1$ can also be written as a 
chain integral. It is given by the Abel-Jacobi 
map
\begin{equation}
A^{(0)}_1(q,x)=\int_{\beta_u} \lambda \ ,
\label{dualopenperiod}
\end{equation}
where $\beta_u$ is now the chain $\beta_u=[u^*,u]$.
Note that the disk amplitude has an integrality structure which may be exhibited by passing to the instanton-corrected coordinates $X=e^U$, $Q = e^{-T}$. Then, it can be written in terms of the open BPS numbers $N_{n,m}^{(0)} \in \IZ$ as follows:
\begin{equation}
A^{(0)}_1(Q,X)=\sum_{n\in \mathbb{N},m\in \mathbb{Z}} 
N^{(0)}_{n,m} {\rm Li}_2(Q^n X^m) \ .
\label{integersuperpotential}
\end{equation}

\subsubsection{Picard-Fuchs equations}

On the Riemann surface it is possible to perform the period 
integrals (\ref{closedmirrormap}), (\ref{openmirrormap}) and \eqref{dualopenperiod} directly. However, in practice
it is simpler to derive Picard-Fuchs equations for general period
integrals, construct a basis of solutions and find linear combinations of the solutions which reproduce the leading behaviour of the period integrals. 

When $M$ is a toric threefold, the Picard-Fuchs operators annihilating the closed periods $T^\alpha$ \eqref{closedmirrormap} can be 
defined in terms of the charge vectors defining $M$ (see \ref{dterm}), as
\begin{equation}
{\cal L}_\alpha=\prod_{Q^\alpha_i>0} \partial_{x_i}-
\prod_{Q^\alpha_i>0} \partial_{x_i}.
\label{generalpicardfuchs}
\end{equation}
The complex structure variables at the points of maximally unipotent monodromy $q^\alpha = \re^{-t^\alpha}$ are  related to the $x_i$ by
\be
q^\alpha=(-1)^{Q^\alpha_0}\prod_{i} x_i^{Q^\alpha_i}. 
\ee
Note that there are in general 
more $x^i$ then  $q^\alpha$ and $\mathbb{C}^*$-scaling symmetries are used to 
reduce to the $q^\alpha$ variables. 

Solutions to (\ref{generalpicardfuchs}) 
are easily constructed using Fr\"obenius method. Defining 
\begin{equation}
w_0({\underline{q}},{\underline{\rho}})=
\sum_{{\underline n}^{\underline{\alpha}}} \frac{1}
{\prod_{i} \Gamma[Q^\alpha_i (n^\alpha+\rho^\alpha)+1]} ((-1)^{Q^\alpha_0} 
q^\alpha)^{n^\alpha},
\end{equation}
then 
\be
X^0=w_0({\underline{q}},{\underline{0}}), \qquad {T}^\alpha=\frac{\partial }{\partial \rho^\alpha}
w_0({\underline{q}},{\underline{\rho}})|_{\underline{\rho}=0}
\ee
  are solutions. 
Higher derivatives 
\be
X^{(\alpha_{i_1}\ldots \alpha_{i_n})}=
\frac{\partial }{\partial \rho^{\alpha_{i_1}}}
\ldots\frac{\partial }{\partial \rho^{\alpha_{i_n}}}
w_0({\underline{q}},{\underline{\rho}})|_{\underline{\rho}=0}
\ee
 also obey the 
recursion imposed by (\ref{generalpicardfuchs}), \emph{i.e.} they fullfill 
(\ref{generalpicardfuchs}) up to finitely many terms. However, only finitely
many linear combinations of the $X^{\alpha_{i_1}\ldots \alpha_{i_n}}$ 
are actual solutions of the Picard-Fuchs system.

Once the solutions $T^\alpha$ to (\ref{generalpicardfuchs}) are given, 
the period integrals (\ref{openmirrormap}) defining the flat open string parameters  
can be simply expressed in terms of them:
\begin{equation}
U = u + \sum_{\alpha=1}^k r_u^\alpha (t^\alpha- T^\alpha)\ .
\label{openmap2}
\end{equation}
Here $r_u^\alpha\in \mathbb{Q}$, and  most of them are zero. 
Note that only the combinations $(t^\alpha- T^\alpha)$ occur, which implies
that the open string variables are invariant under the closed 
string $B$-field shift. 


Note that one can write down an \emph{extended} Picard-Fuchs system, such that not 
only the closed periods but also the open periods \eqref{openmirrormap} and 
\eqref{dualopenperiod} are annihilated by the differential operators \cite{LM,Forbes}. The 
$r^\alpha_u$ are then related to entries in the charge vectors $Q^\alpha_i$ 
in (\ref{dterm}). 
These relations are manifest in the extended Picard-Fuchs system 
and give an easy way to determine the $r^\alpha_u$.

Finally, in the following we will always use the following notation. 
We always denote the flat, instanton corrected coordinates by uppercase 
letters, such as $T$, $U$ and $V$, with their exponentiated counterparts 
$Q = \re^{-T}$, $X = \re^{U}$ and $Y = \re^{V}$. The classical (or uncorrected) 
variables will always be denoted by lowercase lettes $t$, $u$ and $v$, as well as $q = \re^{-t}$, $x=\re^u$ and $y=\re^v$.

\subsubsection{Open phase transitions}

In the example above we have found the open mirror map in a particular parameterization corresponding to outer branes with zero framing. We could have done the same for branes in other phases, but in practice it is easier to simply follow the mirror map through the reparameterizations between different phases in order to obtain the mirror in other phases or framing.

Here we simply write down an explicit example of such calculation.
Let us start with a mirror curve $H(\tx,\ty;q)$ in the  parameterization corresponding 
to outer branes with zero framing. Following \eqref{openmap2}, we can write the open string mirror map, in terms of exponentiated coordinates, as
\be
X = \tx \re^{\Delta_u},
\ee
where
\be
\Delta_u = \sum_{\alpha=1}^k r_u^\alpha (t^\alpha- T^\alpha) .
\ee
Suppose that $\ty$ is not corrected, that is $Y = \ty$, or in the notation above $\Delta_v = 0$.
Consider now the framing transformation
\be
(\tx,\ty) \mapsto (x, y) = (\tx \ty^f, \ty)\, .
\ee
In this case, both the open and closed mirror maps are left unchanged by the framing reparameterization.

Let us now consider a reparameterization corresponding  to
a phase transition to an inner brane phase:
 \be
(\tx,\ty) \mapsto (\tx_i, \ty_i) = \left(\frac{1}{\tx}, \frac{\ty}{\tx}\right) .
\ee
In this case the open mirror maps becomes:
\be
X = \frac{1}{\tx} \re^{\Delta_{u_i}}=\tx_i   \re^{-\Delta_u} ,\qquad Y =\frac{ \ty}{\tx} \re^{\Delta_{v_i}}= \ty_i
\re^{-\Delta_u} \, .
\ee
The fact that $\ty_i$ also gets renormalized in this phase
implies that,  under a  framing reparameterization 
\be
(\tx_i,\ty_i) \mapsto (x_i, y_i) = (\tx_i \ty_i^f, \ty_i) ,
\ee
the open flat coordinates acquire a non-trivial framing dependence:
\begin{eqnarray}\label{genframp}
X&=& \tx_i \ty_i^f  \, \re^{\Delta_{u_i} +f\Delta_{v_i}}= x_i \re^{-(f+1)\Delta_{ u}}\\\nonumber
 Y&=& \ty_i \re^{\Delta_{  v_i}}= y_i \,  \re^{- \Delta_u}.
\end{eqnarray}

\subsubsection{Small radius regions}
\label{phaseflat}

The more interesting case of phase transitions
in the moduli space between patches which require a 
non-trivial symplectic transformation of the closed periods 
can be dealt with as follows.  

On the B-model side, these transitions simply corrrespond to moving 
in the complex structure moduli space beyond the radius of 
convergence of the large radius expansion, or more generally 
from one region of convergence into another. The flat open and 
closed coordinates in all regions are linear combinations of 
the closed periods \eqref{closedmirrormap} and chain integrals \eqref{openmirrormap}, \eqref{dualopenperiod}. The right 
linear combinations that yield the flat open and closed coordinates in this new region can be found using the following requirements:
\begin{itemize}
\item
they should be small enough to be sensible expansion parameters around  
the singularity;
\item the amplitudes should be monodromy invariant when 
expanded in terms of the flat coordinates;
\item the linear combinations giving the flat closed 
coordinates should not involve the chain integrals. 
\end{itemize}

In simple cases this fixes the flat coordinates completely, up to scaling. This was the case, for instance, for the flat closed coordinates of the $\IC^3 / \IZ_3$ orbifold expansion 
of ${\cal O}(-3)\rightarrow \mathbb{P}^2$, which was considered in 
\cite{abk}.

A technical difficulty  is that one has to find local 
expansions of the closed periods and chain integrals at 
various points in the moduli space. For the closed periods this 
can be done by solving the Picard-Fuchs system at the new points 
to obtain a basis of solutions everywhere. For the open periods, one uses the following observations.


First, notice that (\ref{openmap2}) is a 
chain integral, while the $T^\alpha$ are 
periods. Hence there is a linear combination
\begin{equation} 
u_B=u+\sum_{\alpha=1}^k r_u^\alpha t^\alpha\ ,
\label{globalopen}
\end{equation}
which can be written as an elementary function of the global variables $(x,q^\alpha)$. 
Likewise the analytic continuation of $A^{(0)}_1(q,x)$ is trivial 
since it is an elementary function in terms of the global 
variables. Hence, together with the closed string periods, $u_B(q,x)$ 
and $A^{(0)}_1(q,x)$ form a basis for the flat coordinates everywhere in the moduli space.

\subsubsection{The ${\cal O}(-3)\rightarrow \mathbb{P}^2$ geometry}

\label{mirrormapP2}

As an example, let us now discuss the open and closed mirror maps for the ${\cal O}(-3)\rightarrow \mathbb{P}^2$ geometry. Local $\IP^2$ is defined by the charge 
\be
Q=(-3,1,1,1).
\ee 
We start with the closed periods at large radius. Plugging this charge into (\ref{generalpicardfuchs}) and 
changing variables to $q=-\frac{x_2 x_3 x_4}{x_1^3}$, 
we get the Picard-Fuchs differential equation
\begin{equation}
{\cal D}=[\theta_t^2+ 3 q (3 \theta_t+2)(3 \theta_t+1)]\theta,
\end{equation}
where $\theta_t=q \frac{\partial}{\partial q}=\partial_t$. This equation should annihilate the closed periods. 

Clearly $X^0=1$ and 
\be
T:=X^{(t)}=\int_A \lambda=t-\Delta_t(q),
\label{closedflatp2}
\ee
with
\be
\Delta_t(q) = \sum_{n=1}^\infty \frac{(-1)^n}{n} \frac{(3 n)!}{(n!)^3} q^n,
\ee
are solutions. It is easy to check that 
\be
{F}_{T}=
\frac{1}{6} X^{(t,t)}+\frac{1}{6} T+\frac{1}{12}
\ee 
is a third solution, which corresponds to the integral 
$F_{T}=\int_B \lambda$ over the B-cycle. Note that the
particular combination of the Picard-Fuchs solutions giving the B-period is determined by classical 
topological data of the $A$-model geometry. The expression for the flat closed parameter \eqref{closedflatp2} can be inverted to
\be
q=Q+ 6\,Q ^2 + 9\,Q ^3 + 56\,Q ^4 - 300\,Q ^5 + 3942\,Q^6+ \cdots,
\ee
with $q=e^{-t}$ and $Q=e^{-T}$.

\begin{figure}
 \begin{center}
\includegraphics[width=6in]{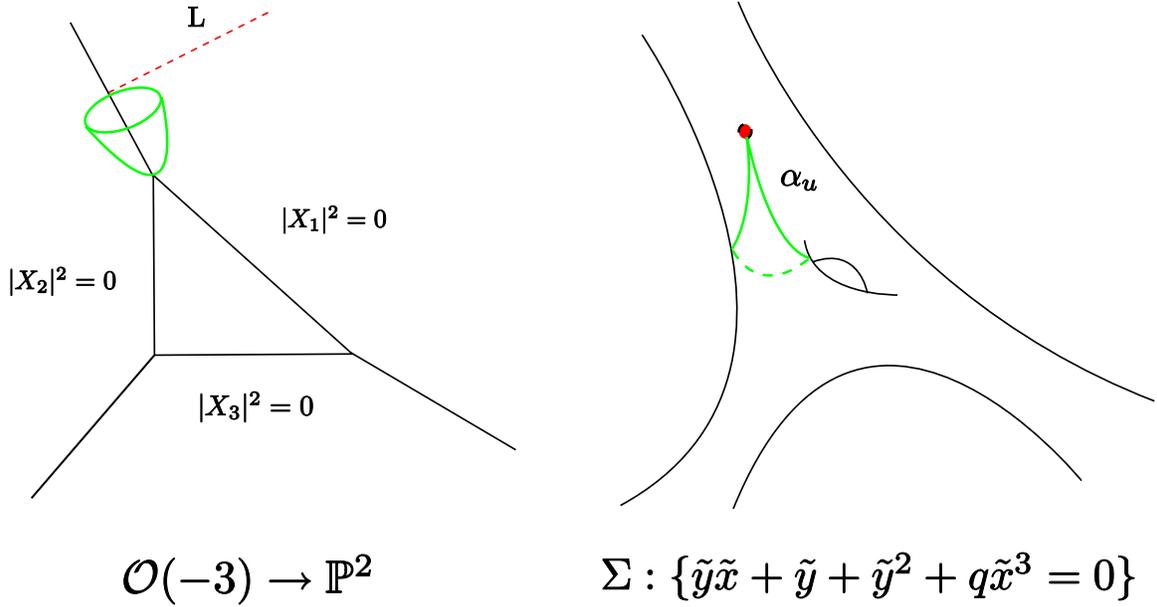}
\end{center}
\caption{Toric base of ${\cal O}(-3)\rightarrow \mathbb{P}^2$ with an
outer brane and the mirror curve with the open cycle 
defining the mirror map.}
\label{figp2}
\end{figure}

We now consider an outer brane in this geometry. Applying (\ref{mirrorcurve}) and 
the discussion in section \ref{openstringphases} we see that 
the parameterization of the mirror curve $H(\tx,\ty;q)$ relevant for 
the outer brane with zero framing is 
\begin{equation}
\label{p2outer}
H(\tx,\ty;q)=\ty^2+\ty+ \ty \tx + q \tx^3=0.
\end{equation}
The derivative of the superpotential is then given by
$\tx \partial_{\tx} A_1^{(0)} =\log(\ty)$, with
\begin{equation}
\label{solouterp2}
\ty=-\frac{1+\tx}{2}-\frac{1}{2}\sqrt{(1+\tx)^2-q \tx^3}.
\end{equation}
The special 
Lagrangian $L$ in the $A$-model becomes a point on the 
Riemann surface; the 
exact domain-wall tension is then given by the period 
integral over the cycle (\ref{openmirrormap}) depicted in 
figure \ref{figp2}. The integral was performed in \cite{akv} and 
yields
\begin{equation}
U= \tilde u -\frac{t-T}{3},
\end{equation}
or
\be
X = \tx \re^{-{1 \over 3} \Delta_t(q)}\, ,
\ee
which defines the open flat coordinate at large radius.
There is no mirror map for $\ty$, that is, $Y= \ty$.

Consider now the framing transformation,
\be
(\tx, \ty) \mapsto (x,y) = (\tx \ty^f, \ty).
\ee
Following this transformations, we get that the open mirror map for framed outer branes is still given by:
\be
X= x \re^{-{1 \over 3}\Delta_t(q)}, \qquad Y = y,
\ee
and its inversion reads
\be
\label{opmap}
x=X\left(1 - 2\,Q + 5\,Q^2 - 32\,Q^3 - 286\,Q^4 +\cdots \right)\, .
\ee

We now move to inner branes. The phase transition from outer branes to inner branes consists in the transformation 
\be
\label{intout}
(\tx,\ty) \mapsto (\tx_i, \ty_i) = \left(\frac{1}{\tx},\frac{\ty}{\tx} \right),
\ee
which gives the curve
\be
\label{p2inner}
H(\tx_i,\ty_i;q)=\ty_i^2 \tx_i + \ty_i \tx_i^2 + \ty_i \tx_i + q,
\ee
parameterizing an inner brane with zero framing. Following the transformation \eqref{intout}, we get that the inner brane mirror map reads
\be
X = \tx_i \re^{{1 \over 3}\Delta_t(q)},\qquad Y = \ty_i \re^{{1 \over 3} \Delta_t(q)}.
\ee
In terms of the framed variables $(x_i, y_i)$, the mirror map becomes
\be
X = x_i \re^{{1 \over 3}(1+f)\Delta_t(q)},\qquad Y = y_i \re^{{1 \over 3}\Delta_t(q)},
\ee
which can be inverted to
\be
\label{inmap}
x_i=X\left( 1 + 2\,\left( 1 + f \right) \,Q + \left( -1 + f + 2\,f^2 \right) \,Q^2 + 
  \frac{2\,\left( 30 + 25\,f - 3\,f^2 + 2\,f^3 \right) \,Q^3}{3} +\cdots\right).
\ee

\section{A new B-model formalism}

In this section we would like to propose a complete method for solving the open and closed B-model topological string on a Calabi-Yau threefold $W$ 
which is the mirror of a toric Calabi-Yau threefold $M$. The method builds on and extends 
the proposal in \cite{mm2}, and it lies entirely in the B-model. It provides in this way a mirror formalism to the A-model topological vertex for toric Calabi-Yau threefolds \cite{topvertex}.

However, our formalism differs from the topological vertex in one crucial aspect. The topological vertex is non-perturbative in $g_s$, the string 
coupling constant, but it is a perturbative expansion in $Q=\re^{-t/\ell_s^2}$ around the large radius point $Q=0$ of the moduli space. In the computation of 
open amplitudes, the vertex is also perturbative in the open moduli $z_i$ appearing for example in (\ref{opena}), and it provides an expansion around $z_i=0$. 
As mentioned earlier, the B-model is perfectly suited for studying the amplitudes at various points in the open/closed moduli space. 
In fact, our formalism provides a recursive method for generating all open and closed amplitudes at any given point in the moduli space. 
Basically, once one knows the disk and the annulus amplitude at this point, one can generate all the other open and closed amplitudes unambiguously. 
In particular, not only can we solve topological string theory at large radius points corresponding to smooth threefolds, 
 but also at other points in the moduli space such as orbifold and conifold points. This is in contrast to the topological vertex, which is defined only for smooth toric Calabi-Yau threefolds.

Our method is recursive in the genus and in the number of holes of the amplitudes, which is reminiscent of the holomorphic anomaly equations of \cite{bcov}. However, a crucial point is that in contrast with the holomorphic anomaly equations, our equations are fully determined, that is, they do not suffer from the holomorphic ambiguities appearing genus by genus when one tries to solve the holomorphic anomaly equations. Our equations are also entirely different in nature from the holomorphic anomaly equations, although it was shown in \cite{emo} that the former imply the latter. More precisely, the 
resulting amplitudes admit a non-holomorphic extension fixed by modular invariance (as in \cite{abk}) which satisfies the holomorphic anomaly equations of \cite{bcov} in the local case. 

The main ingredient that we will make use of is the fact that when $W$ is mirror to a toric Calabi-Yau threefold, most of its geometry is captured by a Riemann surface, which is the mirror curve $\S$ in the notation of the previous section. We will construct recursively an infinite set of meromorphic differentials and invariants living on the mirror curve, and show that the meromorphic differentials correspond to open topological string amplitudes, while the invariants give closed topological string amplitudes. The initial conditions of the recursion are fixed by simple geometric objects associated to the Riemann surface, which encode the information of the disk and the annulus amplitudes.

Our method is in fact a generalization of the formalism proposed by Eynard and Orantin \cite{eo} for solving matrix models. Given a matrix model, one can extract its spectral curve, which is an affine curve in $\IC^2$. Eynard and Orantin used the loop equations of the matrix model to construct recursively an infinite set of meromorphic differentials and invariants on the spectral curve, which give, respectively, the correlation functions and free energies of the matrix model. However, the insight of Eynard and Orantin was that one can construct these objects on \emph{any} affine curve, whether it is the spectral curve of a matrix model or not. The obvious question is then: what do these objects compute in general?

As a first guess, one could try to apply directly Eynard and Orantin to the mirror curve and see what the objects correspond to in topological string theory. However, this would not be correct, since the mirror curve is embedded in $\IC^* \times \IC^*$ rather than $\IC^2$; this is a crucial difference which must be taken into account. But after suitably modifying the formalism such that it applies to curves in $\IC^* \times \IC^*$, it turns out that the objects constructed recursively correspond precisely to the open and closed amplitudes of topological string theory. As argued in \cite{mm2} and as we mentioned in 
the introduction, this is because the formalism of \cite{eo} gives the amplitudes of the chiral boson theory on a ``quantum" Riemann surface constructed in \cite{adkmv}, which should describe as well 
 the B-model on mirrors of toric geometries (once the formalism is suitably modified). 

So let us first start by briefly reviewing the formalism of Eynard and Orantin.

\subsection{The formalism of Eynard and Orantin for matrix models}

Take an affine plane curve
\be
C: \{ \CE(x,y) = 0 \} \subset \IC^2,
\ee
where $\CE(x,y)$ is a polynomial in $(x,y)$. Eynard and Orantin construct recursively an infinite set of invariants $F_g$ of $C$, $g \in \IZ^+$, which they call \emph{genus $g$ free energies}, by analogy with matrix models. The formalism involves taking residues of meromorphic differentials $W_k^{(g)} (p_1, \ldots, p_k)$ on $C$, which are called \emph{genus $g$, $k$ hole correlation functions}. 

\subsubsection{Ingredients}

The recursion process starts with the following ingredients:
\begin{itemize}
\item the ramification points $q_i \in C$ of the projection map $C \to \IC$ onto the $x$-axis, \emph{i.e.}, the points $q_i \in C$ such that ${\partial \CE \over \partial y} (q_i) = 0$. Note that near a ramification point $q_i$ there are two points $q, \bar q \in C$ with the same projection $x(q)=x(\bar q)$;
\item the meromorphic differential 
\be
\label{pheo}
\Phi(p) = y(p) \rd x(p)
\ee
on $C$, which descends from the symplectic form $\rd x \wedge \rd y$ on $\IC^2$;
\item the Bergmann kernel $B(p,q)$ on $C$, which is the unique meromorphic differential with a double pole at $p=q$ with no residue and no other pole, and normalized such that
\be
\oint_{A_I} B(p,q) = 0,
\ee
where $(A_I, B^I)$ is a canonical basis of cycles for $C$.\footnote{Note that the definition of the Bergmann kernel involves a choice of canonical basis of cycles; hence the Bergmann kernel is not invariant under modular transformations --- we will come back to that later.} The Bergmann kernel is related to the prime form $E(p,q)$ by
\be
B(p,q) = \partial_p \partial_q E(p,q).
\ee
We will also need the closely related one-form
\be
\label{deqp}
\rd E_{q} (p)= {1 \over 2} \int_{q}^{\bar q} B(p,\xi),
\ee
which is defined locally near a ramification point $q_i$.
\end{itemize}

For example, if $C$ has genus $0$, its Bergmann kernel is given, in local coordinate $w$, by
\be
\label{bkzero}
B(p,q)= {\rd w (p) \rd w(q) \over (w(p)-w(q))^2 }.
\ee
Note that the Bergmann kernel is defined directly on the Riemann surface, and does not depend on a choice of embedding in $\IC^2$, \emph{i.e.} on the choice of parameterization of the curve. In contrast, by definition the ramification points $q_i$ and the differential $\Phi(p)$ depend on a choice of parameterization of the curve.

Given these ingredients, we can split the recursion process into two steps. First, we need to generate the meromorphic differentials $W_k^{(g)} (p_1, \ldots, p_k)$, and then the invariants $F_g$. 

\subsubsection{Recursion}

\begin{figure}
\leavevmode
 \begin{center}
\epsfxsize=6in
\epsfysize=1.8in
\epsfbox{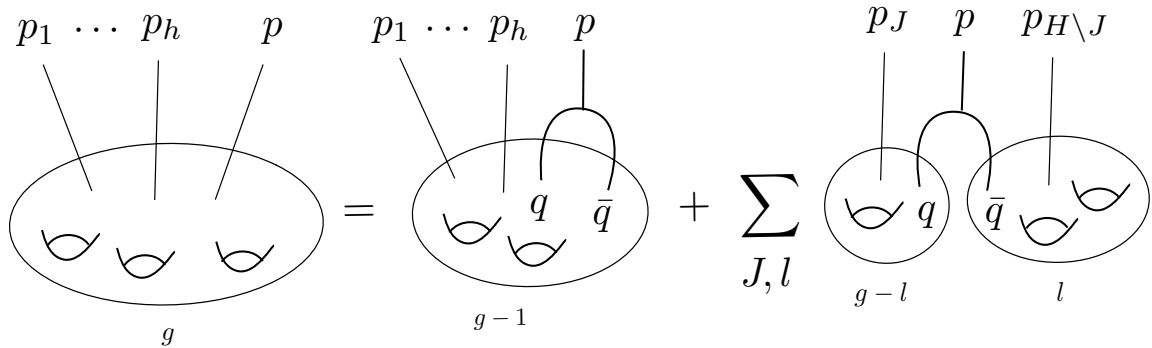}
\end{center}
\caption{A graphic representation of the recursion relation (\ref{rec1}).}
\label{recfig}
\end{figure}

Let $W_h^{(g)} (p_1, \ldots, p_h)$, $g,h \in \IZ^+$, $h \geq 1$, be an infinite sequence of meromorphic differentials on $C$. We first fix
\be
W_1^{(0)}(p_1) = 0, \qquad W_2^{(0)}(p_1,p_2) = B(p_1,p_2),
\ee
and then generate the remaining differentials recursively by taking residues at the ramification points as follows:
\be
\ba
W_{h+1}^{(g)}(p, p_1 \ldots, p_h) &= \sum_{q_i}  \underset{q=q_i}{\rm Res~} {\rd E_{q}(p) \over \Phi(q) - \Phi(\bar q)} \Big ( W^{(g-1)}_{h+2} (q, \bar q, p_1, \ldots, p_{h} )\\
&\qquad +\sum_{l=0}^g \sum_{J\subset H} W^{(g-l)}_{|J|+1}(q, p_J) W^{(l)}_{|H|-|J| +1} (\bar q, p_{H\backslash J}) \Big).
\label{rec1}
\ea
\ee
Here we denoted $H={1, \cdots, h}$, and given any subset 
$J=\{i_1, \cdots, i_j\}\subset H$ we defined $p_J=\{p_{i_1}, \cdots, p_{i_j}\}$. This recursion relation can be represented graphically as in \figref{recfig}.

Now, from these correlation functions we can generate the invariants $F_g$. Let $\phi(p)$ be an arbitrary anti-derivative of $\Phi(p) = y(p) \rd x(p)$; that is, $\rd \phi(p) = \Phi(p)$. We generate an infinite sequence of numbers $F_g$, $g \in \IZ^+$, $g \geq 1$ by
\begin{equation}
F_g = {1 \over 2 - 2 g } \sum_{q_i} \underset{q=q_i}{\rm Res~} \phi(q) W_1^{(g)}(q) .
\label{rec2}
\end{equation}
We refer the reader to \cite{eo} for the formula for the invariant $F_0$, which will not be needed in this paper.

\subsubsection{Symplectic transformations}

As an affine curve in $\IC^2$, the reparameterization group $G_C$ of $C$ is given by 
\be
G_C = SL(2, \IC)  \times \begin{pmatrix} 0 & 1 \\ 1 & 0 \end{pmatrix},
\ee
that is the group of complex $2 \times 2$ matrices with determinant $\pm 1$,
acting on the coordinates $(x,y)$ by
\begin{equation} 
(x,y) \mapsto (a x+ b y, c x + d y), \qquad \left(\begin{array}{cc} a& b\\ c& d\end{array}\right)\in G_C .
\label{sl2c}
\end{equation}
This is the group that preserves the symplectic form $\left| \rd x \wedge \rd y \right|$ on $\IC^2$.

It was shown in \cite{eo} that the free energies $F_g$ constructed as above are \emph{invariants} of the curve $C$, in the sense that they are invariant under the action of $G_C$. However, the correlation functions $W_k^{(g)}(p_1, \ldots, p_k)$ are not invariant under reparameterizations, since they are differentials. 

\subsubsection{Interpretation}

The definition of these objects was inspired by matrix models. When $C$ is the spectral curve of a matrix model, 
the meromorphic differentials $W_k^{(g)}(p_1, \ldots, p_k)$ and the invariants $F_g$ are respectively 
the correlation functions and free energies of the matrix model. To be precise, this is true for 
all free energies with $g \geq 1$, and all correlation functions with $(g,k) \neq (0,1), (0,2)$. 
We refer the reader to \cite{eo} for the definition of the genus $0$ free energy $F_0$. In the case of matrix models, 
the one-hole, genus $0$ correlation function ${\tilde W}_1^{(0)}(p)$ is also known as the 
resolvent and depends on both the potential of the model and the spectral curve, 
\be
{\tilde W}_1^{(0)}(p) ={1\over 2}(V'(p)-y(p)) \rd x (p)
\ee
while the two-hole, genus $0$ correlation function ${\tilde W}_2^{(0)}(p_1,p_2)$ is given by subtracting the double pole from the Bergmann kernel:
\be
{\tilde W}_2^{(0)}(p_1,p_2) = W_2^{(0)}(p_1,p_2) - {\rd p_1 \rd p_2 \over (p_1- p_2)^2} = B(p_1, p_2) - {\rd p_1 \rd p_2 \over (p_1- p_2)^2}.
\ee

\subsection{Our formalism}

As noted earlier, when $W$ is mirror to a toric Calabi-Yau threefold, there is a natural Riemann surface that pops out of the B-model geometry, which is the mirror curve. It is always given by an algebraic curve in $\IC^* \times \IC^*$. Our strategy, extending the proposal in \cite{mm2}, 
will be to apply a recursive process analog to the above to generate free energies and correlation functions living on the mirror curve. 
We will then check extensively that these objects correspond precisely to the open and closed topological string amplitudes.

We start with an algebraic curve
\be
\label{cstarcurve}
\S: \{H(x,y)= 0 \} \in \IC^* \times \IC^*,
\ee
where $H(x,y)$ is a polynomial in $(x,y)$, which are now $\IC^*$-variables. One can think of them as exponentiated variables $(x,y)=(\re^u, \re^v)$, and this 
is how they appeared for example in the derivation of mirror symmetry in \cite{Hori:2000kt}. The only difference with Eynard-Orantin's geometric setup is that our Riemann surfaces are embedded in $\IC^* \times \IC^*$ rather than $\IC^2$. As such, their reparameterization group is the $G_\S$ of \eqref{sl2z} (the group of integral $2 \times 2$ matrices with determinant $\pm 1$), which acts multiplicatively on the $\IC^*$-coordinates of $\S$, rather than the $G_C$ of \eqref{sl2c}. Consequently, we want to modify the recursive formulae such that the free energies $F_g$ constructed from our curve $\S$ are invariant under the action of $G_\S$ given by \eqref{sl2z}. As such, they will be invariants of the Riemann surface $\S$ embedded in $\IC^* \times \IC^*$.

\subsubsection{Ingredients}

The recursion process now starts with the following ingredients:
\begin{itemize}
\item the ramification points $q_i \in \S$ of the projection map $\S \to \IC^*$ onto the $x$-axis, \emph{i.e.}, the points $q_i \in \S$ such that ${\partial H \over \partial y} (q_i) = 0$. Near a ramification point, there is again two points $q, \bar q \in \S$ with the same projection $x(q) = x(\bar q)$;
\item the meromorphic differential 
\be
\label{logth}
\Th(p) = \log y(p) { \rd x(p) \over x(p)}
\ee
on $\S$, which descends from the symplectic form 
\be
{\rd x \over x} \wedge {\rd y \over y}
\ee
on $\IC^* \times \IC^*$. Note that the one-form $\Th(p)$ controls complex structure deformations for the B-model.
\item the Bergmann kernel $B(p,q)$ on $\S$, and the one-form $\rd E_{q}(p)$ defined earlier.
\end{itemize}

The main difference is in the meromorphic differential $\Th(p)$, which differs from the previous differential $\Phi(p)$ because of the symplectic form on $\IC^* \times \IC^*$. Again, both the ramification points $q_i$ and the differential $\Th(p)$ depend on a choice of parameterization for the curve $\S$, while the Bergmann kernel is defined directly on the Riemann surface.

\subsubsection{Recursion}

As before, the recursion process is given in two steps by \eqref{rec1} and \eqref{rec2}; however, we replace the differential $\Phi(p)$ by the new differential $\Th(p)$, to make the formalism suitable for algebraic curves in $\IC^* \times \IC^*$. Accordingly, in \eqref{rec2} $\phi(p)$ is replaced by an arbitrary anti-derivative $\theta(p)$ of $\Th(p)$ as 
defined in \eqref{logth}; that is, $\rd \theta (p) = \Th(p)$.

\subsubsection{Symplectic transformations}

As a curve in $\IC^* \times \IC^*$, the reparameterization group of $\S$ is given by the group $G_\S$ of integral $2 \times 2$ matrices with determinant $\pm 1$, acting on the coordinates $(x,y)$ by
\begin{equation} 
(x,y) \mapsto (x^a y^b, x^c y^d), \qquad \left(\begin{array}{cc} a& b\\ c& d\end{array}\right)\in G_\S . 
\end{equation}
We claim that the $F_g$'s constructed above are invariant under the action of this group, hence are invariants of the mirror curve $\S$. Computationally speaking, a direct consequence of this statement is that we can use the $G_\S$ reparameterizations above to write down the ``simplest" embedding of the Riemann surface in $\IC^* \times \IC^*$, and use this embedding to calculate the free energies. We will use this fact extensively in our computations. Note however again that the correlation functions are not invariant under $G_\S$, which will turn out to be crucial.

\subsubsection{Interpretation}

Suppose now that $\S$ is the mirror curve of a toric Calabi-Yau threefold $M$. Our first claim is:

\begin{description}
\item[1.]{
The free energies $F_g$ constructed above are equal to the A-model closed topological string amplitudes on the mirror threefold $M$, after plugging in the closed mirror map.}
\end{description}

Our second claim is a little bit subtler. Recall that fixing the location and framing of a brane in the A-model corresponds to fixing the $G_\S$ parameterization of the mirror curve $\S$. Hence, the open amplitudes should depend on the parameterization of $\S$. We claim:
\begin{description}
\item[2.]{
The integrated correlation functions $A_k^{(g)} = \int W_k^{(g)}(p_1,\ldots,p_k)$ are equal to the A-model framed open topological string amplitudes on the mirror threefold $M$, after plugging in the closed and open mirror maps.}
\end{description}

This statement means that given a parameterization of $\S$, one can compute the correlation functions, integrate them, plug in the mirror maps, and one obtains precisely the A-model open amplitudes for a brane in the location and framing corresponding to this particular parameterization.

Note that as for matrix models, these claims are true for closed amplitudes with $g \geq 1$, and open amplitudes with $(g,k) \neq (0,1), (0,2)$. The disk amplitude, that is $(g,k) = (0,1)$, is given by \cite{akv, av}
\be
\label{onep}
A_1^{(0)} = \int \Th =\int  \log y { \rd x \over x},
\ee
while the annulus amplitude, $(g,k)=(0,2)$, is given by removing the double pole from the Bergmann kernel:
\be
\label{twop}
A_2^{(0)} = \int \left( B(p_1,p_2) - {\rd p_1 \rd p_2 \over (p_1 - p_2)^2 } \right).
\ee
The one-hole amplitude (\ref{onep}) can be interpreted as the one-point function of a chiral boson living on $\Sigma$ \cite{adkmv}, 
and the Bergmann kernel (\ref{twop}) it just its two-point function \cite{mm}, as expected from the identification of the recursive procedure with the 
theory of the ``quantum" chiral boson on the mirror curve. We will not be concerned with the genus $0$, closed amplitude in this paper.

As a result, we get a complete set of equations, directly in the B-model, that generate unambiguously all genus (framed) open/closed topological string amplitudes for toric Calabi-Yau threefolds. These equations can be understood as some sort of gluing procedure in the B-model, with the building blocks corresponding basically to the disk and the annulus amplitudes. In other words, one only needs to know the disk and the annulus amplitudes, and every other amplitude can be computed exactly using the recursion solution.

Let us finally point out that the approach of \cite{mm2} is a particular case of our more general formalism in the case that the curve can be written as 
\be
\label{hcurve}
y(x)={a(x) + {\sqrt{\sigma (x)}} \over c(x)}, \qquad \sigma(x)=\prod_{i=1}^{2s} (x-x_i). 
\ee
 The choice of $x$, $\bar x$ is as usual a choice of sign 
in the square root, hence the differential (\ref{logth}) is given by 
\be
\label{thetax}
\Theta(x)-\Theta(\bar x)={2\over x} \tanh^{-1} \biggl[ { {\sqrt{\sigma(x)}} \over a(x)} \biggr] \rd x. 
\ee
Therefore, in this particular parameterization, our formalism could be regarded as identical to the formalism of \cite{eo}, albeit for a nonpolynomial 
curve given by 
\be
y_{\rm EO}(x)={1\over x} \tanh^{-1} \biggl[ { {\sqrt{\sigma(x)}} \over a(x)} \biggr].
\ee
 This was the point of view advocated in \cite{mm} (see for example equation (2.17) of that 
paper, where the extra factor of $2$ comes from the contribution of $\bar x$). 
Therefore, the results of \cite{mm} for outer branes with trivial framing are also a consequence 
of our formalism. As it will become clear in the following, curves of the form (\ref{hcurve}) describe 
only a very small class of D-branes, and the right point of view to work in general is precisely the one 
we are developing here. However, and as we will elaborate later on, the curve (\ref{hcurve}) 
is still a useful starting point to compute closed string amplitudes due to symplectic invariance.

\subsection{Computations}

\label{computations}

Let us now spend some time describing how we will carry out calculations to provide various checks of our claims. We also present a more algorithmic version of this formalism that could be applied to compute higher genus/number of holes amplitudes. It could in principle be implemented in a computer code, which we hope to do in the near future.

Most of our calculations will focus on open amplitudes; more precisely, on genus $0$, one-hole (disk), two-hole (annulus) and three-hole amplitudes, and genus $1$, one-hole amplitudes. Let us explain the general idea behind our computations.

From mirror symmetry, we are given an algebraic curve $\S: \{ H(x,y)=0 \}$ in $\IC^* \times \IC^*$, with a $G_\S$ group of reparameterizations acting as in \eqref{sl2z}. These reparameterizations correpond physically to changing the location and framing of the brane. 

\subsubsection{Disk amplitude}

To compute the disk amplitude, which is given by
\be
A_1^{(0)} = \int \Th =\int \log y { d x \over xåÊ},
\ee
all we need to do is to write down $y$ as a function of $x$; that is, we need to solve $H(x,y) =0$ for $y$. This can be done, as a power series in $x$, in any parameterization of $\S$, and after plugging in the mirror map for the open string parameter $x$ in a given parameterization we obtain the framed disk amplitudes for branes ending on any leg of the toric diagram of the mirror manifold. This case was 
studied in detail in \cite{av,akv}. 

\subsubsection{Annulus amplitude}

To compute the annulus amplitude, we need to compute the Bergmann kernel of $\S$. This is trickier. Our strategy, which extends the analysis performed in \cite{mm2}, goes as follows. 

We first use the $G_\S$ reparameterizations to write down the curve $\S$ in a simple form, such as hyperelliptic. This was the case considered in \cite{mm2}. 
Generally, this will correspond physically to a brane ending on an outer leg of the toric diagram, with zero framing (but it does not have to be so). In such a parameterization, there exists explicit formulae to write down the Bergmann kernel of the curve, at least for curves of genus $0$ and $1$. 

For a curve $\S$ of genus $0$, the Bergmann kernel is simply given by
\begin{equation}
B(x_1, x_2) = {\rd y_1 \rd y_2 \over (y_1 - y_2 )^2 },
\end{equation}
where the $y_i$ are defined implicitly in terms of the $x_i$ by $y_i := y(x_i)$, with the function $y(x)$ determined by solving the curve $H(x,y)=0$.

When $\S$ has genus $1$, there is a formula, due to Akemann \cite{akemann}, which expresses the Bergmann kernel of an hyperelliptic curve of genus $1$ in terms of the branch points of the projection map $\S \to \IC^*$ onto the $x$-axis. Let $\lambda_i \in \IC^*$, $i=1, \ldots, 4$ be the four branch points of the projection map. That is, if $q_i \in \S$, $i=1, \ldots, 4$ are the ramification points, then $\lambda_i := x(q_i)$. Then the Bergmann kernel is given by
\begin{multline}
B(x_1,x_2) = {E(k) \over K(k)} {(\lambda_1 - \lambda_3)(\lambda_4 - \lambda_2) \over 4 \sqrt{\prod_{i=1}^4 (x_1 \lambda_i-1)(x_2 \lambda_i - 1)}}  \\
+ {1 \over 4 (x_1 - x_2)^2} \Big ( \sqrt{(x_1 \lambda_1 - 1)(x_1 \lambda_2 -1)(x_2 \lambda_3 -1)(x_2 \lambda_4-1) \over (x_1 \lambda_3 - 1)(x_1 \lambda_4 -1)(x_2 \lambda_1 -1)(x_2 \lambda_2-1)} \\
+ \sqrt{(x_2 \lambda_1 - 1)(x_2 \lambda_2 -1)(x_1 \lambda_3 -1)(x_1 \lambda_4-1) \over (x_2 \lambda_3 - 1)(x_2 \lambda_4 -1)(x_1 \lambda_1 -1)(x_1 \lambda_2-1)}  + 2\Big),
\label{akemann}
\end{multline}
where $K(k)$ and $E(k)$ are elliptic functions of the first and second kind with modulus
\be
k^2 = {(\lambda_1- \lambda_2)(\lambda_3 - \lambda_4) \over (\lambda_1 - \lambda_3) (\lambda_2 - \lambda_4)}.
\label{akemannmod}
\ee
Note that this expression involves an ordering of the branch points, which corresponds to choosing a canonical basis of cycles for the Riemann surface.

Using these explicit formulae, we can integrate the two-point correlation function to get the bare genus $0$, two-hole amplitudes $A_2^{(0)}(x_1, x_2)$ in terms of the open string parameters $x_1$ and $x_2$. We then plug in the open mirror map for that particular parameterization to obtain the open amplitude. 

However, this was done in a particular parameterization, or embedding, which exhibited $\S$ in a simple form, such as hyperelliptic. To obtain the full framed annulus amplitude for branes in other locations, we need to be able to calculate the Bergmann kernel for other parameterizations. But we have seen that the Bergmann kernel is in fact defined directly on the Riemann surface, and does not depend on the particular embedding of the Riemann surface. Hence we can use our result above and simply reparameterize it to obtain the Bergmann kernel of the curve in another parameterization. 

For instance, suppose we are given the Bergmann kernel $B(\tx_1, \tx_2)$ for a curve $\tilde H(\tx,\ty)=0$, and that we reparameterize the curve with the framing transformations $( x,  y) = (\tx \ty^f, \ty)$, $f \in \IZ$ introduced earlier. We obtain a new embedding $H( x,  y) = 0 $ of the Riemann surface. To obtain its Bergmann kernel, we first compute $\tx = \tx ( x)$ as a power series in $ x$, and then reparameterize the Bergmann kernel to get $ B( x_1, x_2) = B( \tx_1( x_1), \tx_2( x_2) )$.

In this way, we are able to compute the bare genus $0$, two-hole amplitude for any framing and brane. To obtain the full result we must then plug in the open mirror map for the open string parameters, in the particular parameterization we are looking at.

\subsubsection{Genus $0$, three-hole amplitude}

To compute the genus $0$, three-hole amplitude, we use the recursion formula \eqref{rec1}. We can also use the simpler formula for the three-point correlation function proved by Eynard and Orantin in \cite{eo}, which reads, for curves embedded in $\IC^* \times \IC^*$:
\begin{equation}
W_3^{(0)}(x_1,x_2,x_3) = \sum_{\lambda_i} \underset{x=\lambda_i}{\rm Res~} B(x,x_1) B(x,x_2) B(x,x_3) { x y(x) \over \rd x \rd y(x)}.
\label{3hgen}
\end{equation}
Using our result for the Bergmann kernel in any parameterization, we can compute the three-point correlation function also in any parameterization. Note however that the branch points $\lambda_i \in \IC^*$ depend on the particular parameterization; hence, when we change parameterization, not only the Bergmann kernel gets reparameterized, but the branch points at which we take residues also change.

Let us now spend a few lines on how to find the ramification points $q_i \in \S$ and the two points $q$ and $\bar q$ satisfying $x(q) = x(\bar q)$ in the neighborhood of a ramification point. First, standard geometry says that the ramification points $q_i$ are defined to be the points satisfying 
\be
{\partial H \over \partial y} (q_i) = 0.
\ee
 The $x$-projection of the ramification points $q_i$ defines the branch points $\lambda_i := x(q_i) \in \IC^*$. The latter can also be found directly as solutions of $\rd x = 0$.

 We will also be interested in determining the branch points of the
 ``framed" curve $H(x, y)$ where $(x, y)=(\tx \,\ty^f,\ty)$; that is, the branch points of the projection on the $x$-axis of the framed curve. These are determined by:
\be
\label{bpm}
\rd x =\rd(\tx \ty^f(\tx))= \ty^{f-1} (\tx) (f \tx \ty'(\tx)+\ty(\tx) )\rd \tx=0.
\ee 
To find all  the branch points $\lambda_i$, one  has to solve  \eqref{bpm}  for all the
 different branches of $\ty(\tx)$. 
 
 We can employ the above equation also to analyze the theory near the branch points:
 given a ramification point $q_i$, and the associated branch point $\lambda_i = x(q_i)$, of the projection on the $x$-axis, we can determine the two points $q, \bar q \in \S$ 
 with the same $x$-projection $x(q) = x(\bar q)$ near $q_i$.
 Define
\be
\label{mirrorpoints}
x(q)=\lambda_i + \zeta, \qquad x(\bar q) =\lambda_i + S(\zeta),
\ee
where
\be
\label{brancheds}
S(\zeta)=-\zeta+ \sum_{k\ge 2} c_k \zeta^k .
\ee
By definition, we have that 
\be
x(q) = (\lambda_i + \zeta) \ty(\lambda_i + \zeta)^f = (\lambda_i + S(\zeta)) \ty(\lambda_i + S(\zeta))^f = x(\bar q), 
\ee
which can be used to determine $S(\zeta)$. At the first orders, we get
\be
\ba
c_2&=-\frac{2\,\left( -1 + f^2 \right) \,\ty(\lambda_i) + 
    f^2\,{\lambda_i}^2\,\left( 3\,\ty''(\lambda_i) + 
       f\,\lambda_i\,\ty^{(3)}(\lambda_i) \right) }{3\,f\,\lambda_i\,
    \left( \left( -1 - f \right) \,\ty(\lambda_i) + 
      f^2\,{\lambda_i}^2\,\ty''(\lambda_i) \right) }, \\
      c_3&=-\frac{{\left( 2\,\left( -1 + f^2 \right) \,\ty(\lambda_i) + 
       f^2\,{\lambda_i}^2\,\left( 3\,\ty''(\lambda_i) + 
          f\,\lambda_i\,\ty^{(3)}(\lambda_i) \right)  \right) }^2}{9\,
    f^2\,{\lambda_i}^2\,{\left( \left( 1 + f \right) \,\ty(\lambda_i) - 
        f^2\,{\lambda_i}^2\,\ty''(\lambda_i) \right) }^2}.
      \ea
      \ee

\subsubsection{Genus $1$, one-hole amplitude}

To compute the genus $1$, one-hole amplitude, we also use the recursion formula \eqref{rec1}, with the ramification points and the Bergmann 
kernel corresponding to the chosen parameterization. The general formula for $W_1^{(1)}(q)$ is
\be
\label{wone}
W_1^{(1)}(p) =\sum_{q_i} {\rm Res}_{q=q_i} {\rd E_{q} (p) \over \Phi(q)-\Phi(\bar q)} B(q,\bar q).
\ee

\subsubsection{Higher amplitudes}

Computations at higher $g,h$ can be readily made in this formalism, although they are more complicated. When the algebraic curve is of genus zero, the 
computations are straightforward, but they become more involved as soon as the curve has higher genus. Some simplifications arise however when 
the curve is of the form (\ref{hcurve}) and the differential $\Th(x)$ is of the form (\ref{thetax}), since in this case one can adapt the detailed 
results of \cite{eynard} to our context (see also \cite{bertoldi} for examples of detailed computations). We will refer to this case as the hyperelliptic case, 
since the underlying geometry is that of a hyperelliptic curve. 
Let us briefly review this formalism, following \cite{eynard} closely, in order to sketch how to compute systematically higher amplitudes. We first write 
\be
\label{thxh}
\Th(x)-\Th(\bar x)=2 M(x) {\sqrt {\sigma (x)}} \rd x, 
\ee
where $\sigma(x)$ is defined in (\ref{hcurve}) and $M(x)$ is called the moment function. 
In the formalism of \cite{eo} applied to conventional matrix models, $M(x)$ is a polynomial. In our formalism for 
mirrors of toric geometries, in the 
parametrization of the curve given in (\ref{hcurve}), $M(x)$ is given by
\be
M(x)={1\over x {\sqrt {\sigma (x)}}} \tanh^{-1} \biggl[ { {\sqrt{\sigma(x)}} \over a(x)} \biggr],
\ee
which is the moment function considered in \cite{mm2} (again, up to a factor of $2$ which comes from (\ref{thxh}) and in \cite{mm2} 
is reabsorbed in the definition of $M(x)$). When $\Th(x)$ is of the form (\ref{thxh}) we are effectively working on the 
hyperelliptic curve of genus $s-1$
\be
\label{hc}
y^2(x) =\sigma(x),
\ee
with ramification points at $x=x_i$, $i=1, \cdots, 2s$. We define the $A_j$ cycle of this curve 
as the cycle around the cut
\be
(x_{2j-1}, x_{2j}), \qquad j=1, \cdots, s-1.
\ee
There exists a unique set of $s-1$ polynomials of degree $s-2$, denoted by $L_j(x)$,
such that the differentials 
\be
\omega_j ={1\over 2\pi \ri} {L_j(x)\over \ssq{x}}\rd x
\ee
satisfy
\be
\oint_{A_j}  \omega_i = \delta_{ij}, \qquad i,j=1, \cdots, s-1.
\ee
The $\omega_i$s 
are called normalized holomorphic differentials. The one-form (\ref{deqp}) can then be written as \cite{eynard}
\be
\label{CjLambda}
\rd E_{x'}(x)={1\over 2} {\ssq{x'}\over \ssq{x}}\,\left(
{1\over x-x'}-\sum_{j=1}^{s-1} C_j(x')L_j(x)
\right)\, \rd x
\ee
where
\be\label{defCj}
C_j(x'):={1\over 2\pi \ri }\,\oint_{A_j} {\rd {x}\over {\sqrt{\sigma(x)}}}\,{1\over x-x'}
\ee
In this formula, it is assumed that $x'$ lies outside the contours $A_j$.
One has to be careful when $x'$ approaches some branch point $x_j$.
When $x'$ lies inside the contour $A_j$, then one has:
\be
\label{regc}
C^{\rm reg}_l(x')+{\delta_{lj}\over \ssq{x'}}={1\over 2\pi \ri}\,\oint_{A_j} {\rd {x}\over \ssq{x}}{1\over x-x'}
\ee
which is analytic in $x'$ when $x'$ approaches $x_{2j-1}$ or $x_{2j}$. 
The Bergmann kernel is then given by:
\be\label{BergmanCjdS}
B(x,x') =  \rd x'\,{\rd\over \rd x'}
\left({\rd x\over 2(x-x')}+\rd E_{x'}(x)\right),
\ee
and it can be equivalently written as
\be
\label{twopoint}
\ba
{ B(p,q) \over \rd p \rd q} =& {1\over 2(p-q)^2}
+ {\sigma(p)\over 2(p-q)^2\ssq{p}\ssq{q}} \\
& - {\sigma'(p)\over 4(p-q)\ssq{p}\ssq{q}}
+ {A(p,q)\over 4\ssq{p}\ssq{q}}
\ea
\ee
where $A(p,q)$ is a polynomial. In the elliptic case $s=2$, there is one single integral $C_1(p)$ to 
compute, and one can find very explicit expressions in terms of elliptic integrals:
\be
\label{genintegral}
\ba
C_1(p)&= {2\over \pi (p-x_3) (p-x_2) {\sqrt{(x_1-x_3)(x_2-x_4)}}} 
\biggl[ (x_2-x_3) \Pi(n_4, k) + (p-x_2) K(k)\biggr], \\
C^{\rm reg}_1(p)&=  {2\over \pi (p-x_3) (p-x_2) {\sqrt{(x_1-x_3)(x_2-x_4)}}} 
\biggl[ (x_3-x_2) \Pi(n_1, k) + (p-x_3) K(k)\biggr]
\ea
\ee
where 
\be
k^2 ={(x_1-x_2)(x_3-x_4)\over(x_1-x_3)(x_2-x_4)}, \qquad n_4= {(x_2-x_1)(p-x_3) \over (x_3-x_1)(p-x_2)}, \qquad n_1= {(x_4-x_3)(p-x_2) \over (x_4-x_2)(p-x_3)},
\ee
$\Pi(n, k)$ is the elliptic integral of the third kind,
\be
\Pi(n, k)=\int_0^1 {\rd t\over (1-n t^2) {\sqrt {(1-t^2)(1-k^2 t^2)}}}
\ee
and $K(k)$ is the standard elliptic integral of the second kind. 

With these ingredients one can compute the residues as required in (\ref{rec1}). It is easy to see that $\rd E_{q}(p)/y(q)$, as a function of $q$, 
has a pole at $q=p$ but no pole at the branchpoints. It is then easy to see that all residues appearing in (\ref{rec1}) will be linear combinations of the following 
kernel differentials
\be
\label{kd}
\chi_i^{(n)}(p)={\rm Res}_{q=x_i} \biggl(  {\rd E_q (p) \over y(q)} {1\over (q-x_i)^n} \biggr)
\ee
which are explicitly given by
\be
\label{kdexp}
\chi_i^{(n)}(p)={1\over  (n-1)!} {1\over  {\sqrt {\sigma(p)}}} {\rd^{n-1} \over \rd q^{n-1}} \Biggl[{1\over 2 M(q) }\biggl( {1\over p-q} -\sum_{j=1}^{s-1} L_j(p) C_j(q)\biggl) \Biggr]_{q=x_i}.
\ee
Notice that in order to compute the kernel differentials, the only nontrivial objects to compute are $\rd^k C_j / \rd q^k$. For a curve 
of genus one, they can be evaluated from the explicit 
expressions in (\ref{genintegral}). In order to compute the residues involved in (\ref{rec1}), one has to take into account that the residues 
around branchpoints in terms of a local coordinate as in (\ref{rec1}) 
are twice the residues around $x=x_i$ in the $x$ plane \cite{eynard}. One then finds, for example, 
\be
\ba
W_0(p_1, p_2, p_3)&={1\over 2} \sum_{i=1}^{2s} M^2(x_i) \sigma'(x_i) \chi^{(1)}_i(p_1) \chi^{(1)}_i(p_2) \chi^{(1)}_i(p_3),\\
W_1(p)&={1\over 16} \sum_{i=1}^{2s} \chi^{(2)}_i(p) +{1\over 8} \sum_{i=1}^{2s} \biggl( 2{A(x_i, x_i) 
\over \sigma'(x_i)} -\sum_{j\not=i} {1\over x_i -x_j} \biggr) \chi_i^{(1)}(p),
\ea
\ee
where $A(p,q)$ is the polynomial in (\ref{twopoint}). 

Therefore, in the hyperelliptic case, 
when $\Th(x) -\Th(\bar x)$ can be written as in (\ref{thxh}), the computation of the amplitudes can be done by residue calculus and the 
only part of the calculation which is not straighforward is the evaluation of the integrals (\ref{defCj}), (\ref{regc}). In the elliptic case, they reduce 
to elliptic functions, as we saw in (\ref{genintegral}). 
In the general case one can evaluate the integrals in terms of suitable generalizations of elliptic functions.

\subsection{Moving in the moduli space}

\label{movingBmodel}

In section \ref{modulispace} we discussed in some detail phase transitions in the open/closed string moduli space. We explained why the B-model was perfectly suited for studying such transitions. We now have a formalism, entirely in the B-model, that generates unambiguously all open/closed amplitudes for toric Calabi-Yau threefolds. An obvious application is then to use this formalism to study both open and closed phase transitions, which cannot be studied with A-model formalisms such as the topological vertex.

Recall that the ingredients in our formalism consists in a choice of projection $\S \to \IC^*$ (or equivalently a choice of parameterization of $\S$), a differential $\Th(p)$ corresponding to the disk amplitude, and the Bergmann kernel $B(p,q)$ of the curve --- which yields the annulus amplitude. Note that once the parameterization is chosen, the one-form $\Th(p)$ is canonically defined to be
\be
\Th(p) = \log y(p) {\rd x(p) \over x(p)}.
\ee
Hence $\Th(p)$ really only depends on the choice of parameterization.

We have seen that changing the parameterization of the curve $\S$ corresponds to changing the location and framing of the branes, that is, moving in the open moduli space. This is the mildest type of transition that was considered in section \ref{moving}. Since the Bergmann kernel is really define on the Riemann surface, it can simply be reparameterized, and open phase transitions are rather easy to deal with. As explained in section \ref{moving}, this is because the amplitudes are simply rational functions of the open string moduli, which we see explicitly in our formalism.

The more interesting types of transitions are thus the transitions between different patches which require non-trivial ${\rm Sp}(2 g, \IC)$ transformation of the periods. The only ingredient that is modified by these transitions in the closed string moduli space is the Bergmann kernel, since its definition involves a choice of canonical basis of cycles, which corresponds to a choice of periods.

Modular properties of the Bergmann kernel have been studied in detail in \cite{eo, emo}. Under modular transformations, the Bergmann kernel transforms with a shift as follows:
\be
B(p,q) \mapsto B(p,q) - 2 \pi i \omega(p) (C \tau + D)^{-1} C \omega(q),
\ee
with
\be
\begin{pmatrix} A & B \\ C & D \end{pmatrix} \in {\rm Sp}(2g, \IZ),
\ee
and $\tau$ is the period matrix. Here, $\omega(p)$ is the holomorphic differentials put in vector form. In a sense, the Bergmann kernel is an open analog --- since it is a differential in the open string moduli --- of the second Eisenstein series $E_2(\tau)$, which also transforms with a shift under $SL(2,\IZ)$ transformations and generates the ring of quasi-modular forms.

The key point here is that we know how the Bergmann kernel transforms under phase transitions in the closed string moduli space. Hence not only can we use our formalism to generate the amplitudes anywhere in the open moduli space, but also in the full open/closed moduli space. This means that in principle, we can generate open and closed amplitudes for target spaces such as conifolds or orbifolds. We will explore this avenue further in section \ref{orbifold}.

To end this section, let us be a little more precise. In this paper we will only consider $S$-duality transformations for curves of genus $1$, which  exchange the A- and the B-cycles. More precisely, the $S$-duality transformation acts on the basis of periods by
\be
\begin{pmatrix} 0 & -1 \\ 1 &0 \end{pmatrix} \in SL(2,\IZ).
\ee
When the curve has genus $1$, we can use Akemann's expression \eqref{akemann} to compute the Bergmann kernel. This expression depends on the branch points $\lambda_i$, $i=1,\ldots, 4$, and the choice of canonical basis (or periods) is encoded in the choice of ordering of the branch points. In terms of the elliptic modulus $k^2$, the $S$-duality transformation is given by
\be
k^2 \mapsto  1-k^2 .
\ee
Using the explicit expression for the modulus in terms of the branch points \eqref{akemannmod}, we see that the $S$-transformation is given by exchanging the two branch points $\lambda_2$ and $\lambda_4$. In other words, an $S$-duality transformation corresponds to the two cuts meeting at one point and then splitting again.
Therefore, to determine the shifted Bergmann kernel after an $S$-duality transformation, we only need to use Akemann's expression \eqref{akemann} again, but with $\lambda_2$ and $\lambda_4$ exchanged. Using this new Bergmann kernel we can generate all open and closed amplitudes after the phase transition corresponding to the $S$-duality transformation. 

We will exemplify this procedure in section \ref{orbifold}, where we use an $S$-duality phase transition to compute open and closed amplitudes at the point in the moduli space of local $\IP^1 \times \IP^1$ where the two $\IP^1$'s shrink to zero size. Using large $N$ duality, we can compare the resulting amplitudes with the expectation values of the framed unknot in Chern-Simons theory on lens spaces, and we find perfect agreement.

\section{Genus $0$ examples}

In this section we study two toric Calabi-Yau threefolds, $\IC^3$ and the resolved conifold, for which the mirror curve has genus $0$.

\subsection{The vertex}

Our first example is the simplest toric Calabi-Yau threefold, $M = \IC^3$. The mirror curve $\S$ is $\IP^1$ with three holes, and can be written algebraically as
\begin{equation}
\tilde H(\tx,\ty) = \tx+\ty+1 =0,
\end{equation}
with $\tx,\ty \in \IC^*$.\footnote{In the following, tilde variables will always denote a curve in zero framing, while plain variables will denote a framed curve.}

This parameterization corresponds to a brane ending on one of the three outer legs of the toric diagram, with zero framing (in standard conventions). The open mirror map, in this parameterization, is given simply by $(X,Y) = (-\tx,-\ty)$.

\subsubsection{Framing}

\begin{figure}
\begin{center}
\includegraphics[width=5in]{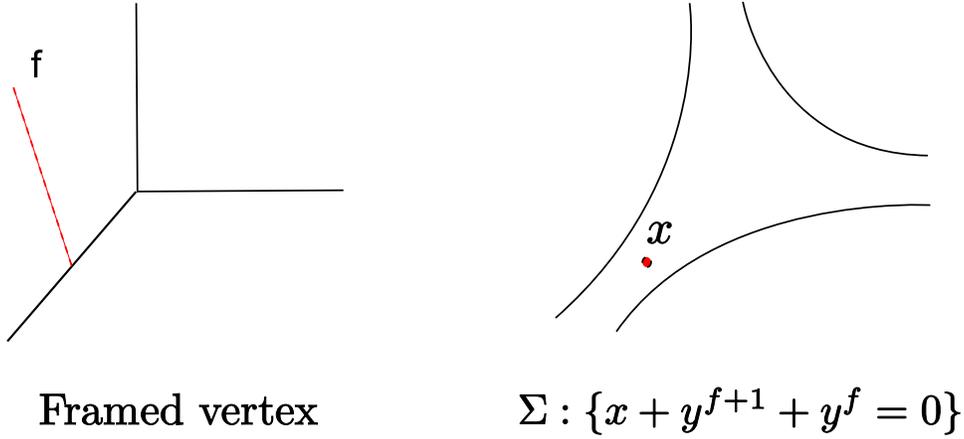}
\end{center}
\caption{The framed vertex and its mirror curve.}
\label{f:vertex}
\end{figure}

The framing transformation is given by
\be
(\tx,\ty) \mapsto (x, y) = (\tx \ty^f, \ty),
\ee
where $x$ is the framed bare open string parameter. From the transformation above, the open mirror map is now given by $(X,Y) = ((-1)^{f+1} x, - y)$. Under this reparameterization the mirror curve becomes
\begin{equation}
H(x, y) = x + y^{f+1} + y^f = 0,
\label{fvertex}
\end{equation}
which is a branched cover of $\IC^*$. The framed vertex and its mirror curve are shown in figure \ref{f:vertex}.

\subsubsection{Disk amplitude}

The bare framed disk amplitude is given by
\be
A_1^{(0)} (x) = \int \log y(x) {\rd x \over x}. 
\ee
Thus, we need to find $y = y(x)$. We can solve \eqref{fvertex} for $y$ as a power series of $x$, by using for example Lagrange inversion, and we get
\begin{equation}
\ba
y(x) &=-1+\sum_{n=1}^{\infty} (-1)^{n(f+1)} {(n f + n -2)! \over (n f -1)!} {x^n \over n!}\\
&=
-1 - (-1)^f x + f x^2 -{(-1)^f \over 2} (f + 3 f^2) x^3 + \ldots
\label{expvert}
\ea
\end{equation}

Plugging in the map $x = - (-1)^f X$, we thus get
\begin{multline}
A_1^{(0)}(X) =  - \Big( X + {1 \over 4} (1+2f) X^2 +{1 \over 18} (2 + 9f + 9f^2) X^3 \\
+ {1 \over 48} (3 + 22 f + 48 f^2 + 32 f^3) X^4 + \ldots \Big),
\end{multline}
up to an irrelevant constant of integration. This is precisely the result that is obtained on the A-model using the topological vertex.

\subsubsection{Annulus amplitude}

To compute the annulus amplitude we must compute the Bergmann kernel of the curve \eqref{fvertex} in the bare open string parameters $x_1$ and $x_2$. 

Let us first work in the zero framing parameterization. Since $\S$ has genus $0$, at zero framing the Bergmann kernel is simply given by
\be
B(\tx_1, \tx_2) = {\rd \ty_1 \rd \ty_2 \over ( \ty_1 - \ty_2)^2},
\ee
where the $\ty_i$ are defined implicitly in terms of the $\tx_i$ by $\ty_i := \ty(\tx_i)$, with $\ty(\tx)$ obtained by solving $\tilde H(\tx,\ty) = 0$, that is $\ty(\tx) = -1 -\tx$.

But the framing transformation sets $y_1 = \ty_1$, $y_2 = \ty_2$, hence we can reparameterize the Bergmann kernel and obtain immediately that
\be
B(x_1, x_2) = {\rd y_1 \rd y_2 \over ( y_1 - y_2)^2},
\ee
where now the $y_i$ are defined implicitly in terms of the $x_i$ by $y_i:=y(x_i)$, with the function $y(x)$ given by \eqref{expvert}.

The bare two-hole amplitude is given by removing the double pole and integrating:
\begin{align}
A_2^{(0)}(x_1, x_2) =& \int \left( B(x_1, x_2) - {\rd x_1 \rd x_2 \over (x_1 - x_2)^2} \right)\notag\\
=& \log (- y_1(x_1) + y_2( x_2) ) - \log(-x_1 + x_2).
\end{align}
Using the expansion \eqref{expvert} and the open mirror map $X_1 = -(-1)^f x_1$, $X_2 = -(-1)^f x_2$, we obtain
\begin{multline}
A_2^{(0)}(X_1,X_2) = {1 \over 2} f (f+1) X_1 X_2 + {1 \over 3} f (1+3f+2f^2) (X_1^2 X_2 + X_1 X_2^2) \\
+ {1 \over 4} f (1+f) (1+2f)^2 X_1^2 X_2^2 + {1 \over 8} f(2+11f+18f^2+f^3) (X_1^3 X_2 + X_1 X_2^3) + \ldots
\end{multline}
up to irrelevant constants of integration; this matches again the topological vertex result.

\subsubsection{Three-hole amplitude}

To compute $A_3^{(0)}$, the additional ingredients needed are the ramification points of the projection map $\S \to \IC^*$ onto the $x$-axis for the framed curve \eqref{fvertex}. Solving
\be
{\partial  H \over \partial y} =0,
\ee
we find only one ramification point $q_1$ at $y(q_1) = - {f \over f+1}$. Denote by $\lambda_1$ the associated branchpoint, which is given by the $x$-projection of $q_1$, that is $\lambda_1 = x(q_1)$.

The amplitude thus becomes
\begin{align}
A_3^{(0)} (x_1, x_2, x_3) =& \int \underset{x = \lambda_1}{\rm Res~} B(x,x_1) B(x,x_2) B(x,x_3) {x y(x) \over \rd x \rd x {\rd y \over \rd x} } \notag\\
=& \int \underset{y = -{f \over f+1}}{\rm Res~} {x(y) y \rd y \rd y_1 (x_1) \rd y_2(x_2) d y_3(x_3) \over (y-y_1(x_1))^2 (y-y_2(x_2))^2 (y - y_3 (x_3) )^2} \left({\rd x \over \rd y}\right)^{-1}.
\end{align}
Since $x = - y^f (y +1)$, we compute easily that
\be
\left({\rd x \over \rd y}\right)^{-1} = - {1 \over y^{f-1} (f+y (f+1) )},
\ee
which has a simple pole at $y = -{f \over f+1}$. Taking the residue and integrating, we get
\begin{align}
A_3^{(0)} (x_1, x_2, x_3) =& - \int {f^2 (f+1)^2} \prod_{i=1}^3 {\rd y_i (x_i) \over (f + (f+1) y_i(x_i) )^2 } \notag\\
=& {f^2 \over f+1} \prod_{i=1}^3 {1 \over f + (f+1) y_i(x_i)}.
\end{align}
Plugging in the expansion \eqref{expvert} and the open mirror map, we finally obtain
\begin{multline}
A_3^{(0)} (X_1, X_2, X_3) = - \Big ( f^2(1+f)^2 X_1 X_2 X_3 + f^2 (1+f)^2 (1+2f) (X_1^2 X_2 X_3 + \text{perms}) \\
+ {1 \over 2} f^2 (1+f)^2 (2+9f + 9f^2) (X_1^3 X_2 X_3 + \text{perms}) +  f^2 (1 +3 f + 2f^2)^2 (X_1^2 X_2^2 X_3 + \text{perms}) + \ldots \Big),\notag
\end{multline}
which is again in agreement with vertex computations.

\subsubsection{The genus one, one hole amplitude}

In the computation of $A_1^{(1)}(X)$ we need some extra ingredients, besides the ones that we have already considered. 
For a curve of genus zero, 
\be
\rd E_{q}(p)= {1\over 2} \rd y(p) \biggl[ {1\over y(p)-y(q)} - {1\over y(p)-y(\bar q)}\biggr],
\ee
where $y$ is a local coordinate.
To compute (\ref{wone}) in this example, we need $\bar q$ near the ramification point $q_1$ located at $y(q_1)=-{f \over 1+f}$. Following the general discussion in section \ref{computations}, we write 
\be
y(q)=-{f\over 1+f}+\zeta, \qquad y(\bar q) = -{f\over 1+f}+S(\zeta).
\ee
By definition,
\be
\label{barq}
x(q) = - y(q)^f (y(q)+1) = - y(\bar q)^f (y(\bar q) + 1) = x(\bar q),
\ee
which we can use to solve for $S(\zeta)$, which has the structure presented in \eqref{brancheds}. Its power series expansion can be easily determined, and the first few terms are
\be
S(\zeta) = -\zeta +\frac{2\,\left( -1 + f^2 \right) \, \zeta^2}{3\,f} - \frac{4\,{\left( -1 + f^2 \right) }^2\, \zeta^3}{9\,f^2} +\CO(\zeta^4). 
\ee
We now compute (\ref{wone}) by using $\zeta$ as a local coordinate near the branchpoint. We need, 
\be
B(q,\bar q)={(\rd \zeta)^2 \over (\zeta -S(\zeta))^2} S'(\zeta),
\ee
as well as 
\be
\Phi(q)-\Phi(\bar q) =\biggl(\log \Bigl( -{f\over 1+f} +\zeta\Bigr)-\log \Bigl( -{f\over 1+f} +S(\zeta)\Bigr)\biggr) {\rd  x \over \rd \zeta} \rd \zeta.
\ee
The residue in (\ref{wone}) is easily evaluated, and we only need the expansion of $S(\zeta)$ up to third order. 
One finds, 
\be
W^{(1)}_1(y)=\frac{ (1+f)^4 y^2 +2f (1+f) (2+f+f^2)y +f^4 }{24\,{\left( f\,\left( 1 + p \right)  + p \right) }^4} \rd y,
\ee
After integration and expanding in $X$, we obtain 
\be
\ba
A^{(1)}_1(y)&=-\frac{ X}{24}+\frac{(1+ 2f)(f^2 +f -1) X^2}{12} \\ &+
  \frac{ (1+ 3f)(2+3f)(-1+2f+ 2f^2)\, X^3}{16} +\CO(X^4),
 \ea
  \ee
which is in perfect agreement with the $g=1$ piece of the exact formula in $g_s$ (but 
perturbative in $X$) obtained from the topological vertex, 
\be
A_1(y,g_s) =\sum_{g=0}^{\infty} A^{(g)}_1(y) g_s^{2g-1} =\sum_{m=0}^{\infty} {[m f + m-1]! \over m [m]! [mf]!}  (-1)^{mf} X^{m+1}, 
\ee
where $[n]$ denotes the $q$-number with parameter $q=\re^{g_s}$. 

To end this section, we mention that the framed vertex results can be written down in a nice way in terms of Hodge integrals, using the Mari\~no-Vafa formula \cite{mv}. The recursion relations proposed in this paper induce new recursion relations for the Hodge integrals. In turn, using the well known relation between the framed vertex geometry and Hurwitz numbers, one can obtain a full recursion solution for Hurwitz numbers. This is a nice mathematical consequence of the formalism proposed in this paper, which is studied in \cite{BM}.
 
\subsubsection{Framed vertex in two legs}

\begin{figure}
\begin{center}
\includegraphics[width=2in]{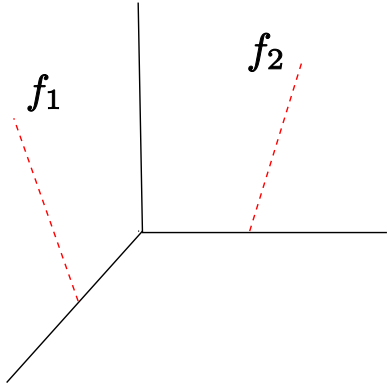}
\end{center}
\caption{The framed vertex in two legs.}
\label{f:vertex2}
\end{figure}

So far we assumed that all the branes ended on the same leg of the toric diagram of $\IC^3$ (the vertex). However, when there are more than one hole, one can consider the case where there is one brane in one leg of the vertex and another brane in another leg; this is shown in figure \ref{f:vertex2}. Let us now compute the annulus amplitude for two branes in two different legs. The strategy goes as usual: we start with the Bergmann kernel for two branes with zero framing in the same leg, and then reparameterize the Bergmann kernel to obtain two framed branes in different legs. 

To do so, we need to find the expansion $y_1 = y_1(x_1)$ for a framed brane in one leg, which we found already in \eqref{expvert}, but also $y_2 = y_2(x'_2)$, where $x'_2$ now corresponds to the open string parameter of a framed brane in a different leg. That is, we need to be able to relate the curves in the two different legs.

As explained in section \ref{modulispace}, the phase transformation for moving from one leg of the toric diagram to another, at zero framing, reads:
\be
(\tx,\ty) \mapsto (\tx', \ty') = (\tx^{-1}, \tx^{-1} \ty).
\ee
 Now the framing transformation in this new leg reads
\be
(\tx', \ty') \mapsto (x',y') = (\tx' (\ty')^f, \ty'),
\ee
where $x'$ and $y'$ now correspond to framed parameters in the new leg. Combining these two transformations we get
\be
(\tx,\ty) \mapsto (x', y') = (\tx^{-1-f} \ty^f, \tx^{-1} \ty),
\ee
Inversely, we have that
\be
(\tx,\ty) = ((x')^{-1} (y')^f, (x')^{-1} (y')^{f+1} ).
\ee
Under this reparameterization the curve becomes
\be
x' + (y')^f + (y')^{f+1} = 0,
\ee
which is the same curve as before! Indeed, for the framed vertex, by symmetry changing the leg does not change the amplitudes. 

So we know $y'(x')$ which is \eqref{expvert} as before. However, what we really want in order to reparameterize the Bergmann kernel is $\ty = \ty(x')$. Using the transformation above, we know that $\ty = \ty(x') = (x')^{-1} (y'(x'))^{f+1}$. As a power series, we get
\begin{equation}
\ty = -(-1)^f {1 \over x'} - (1+f) + {(-1)^f \over 2} f (1+f) x' +\ldots
\label{expvertother}
\end{equation}

Using these results, we can reparameterize the Bergmann kernel to get the framed annulus amplitude in two different legs. For the first open string parameter, we reparameterize using $\ty_1 = y_1 = y_1 (x_1)$ given by \eqref{expvert}, and for the second open string parameter we use \eqref{expvertother} to get $\ty_2 = \ty_2 (x'_2)$. The mirror map for the first parameter is $X_1 = - (-1)^{f_1} x_1$, while for the second parameter from the transformations above we get the mirror map $X_2 = - x'_2$. Removing the double pole and integrating as usual, we get\footnote{Note that here we have two framings $f_1$ and $f_2$ corresponding to the two different branes.}
\begin{multline}
A_2^{(0)}(X_1, X_2) = - (-1)^{f_2} X_1 X_2 -  f_2 X_1 X_2^2 -{(-1)^{f_2} \over 2}(f_2(1+3 f_2) ) X_1 X_2^3 \\
- {1 \over 2} (1 + 2 f_1 f_2) X_1^2 X_2^2 - {(-1)^{f_2} \over 2} f_2 (2+f_1 + 3f_2 f_1) X_1^2 X_2^3 + \ldots
\end{multline}
which again is in agreement with the vertex result.\footnote{More precisely, to get the topological vertex result we need to redefine $f_1 \mapsto -f_1 -1$, which is just a redefinition of what we mean by zero framing.}

\subsection{The resolved conifold}

\begin{figure}
\begin{center}
\includegraphics[width=5in]{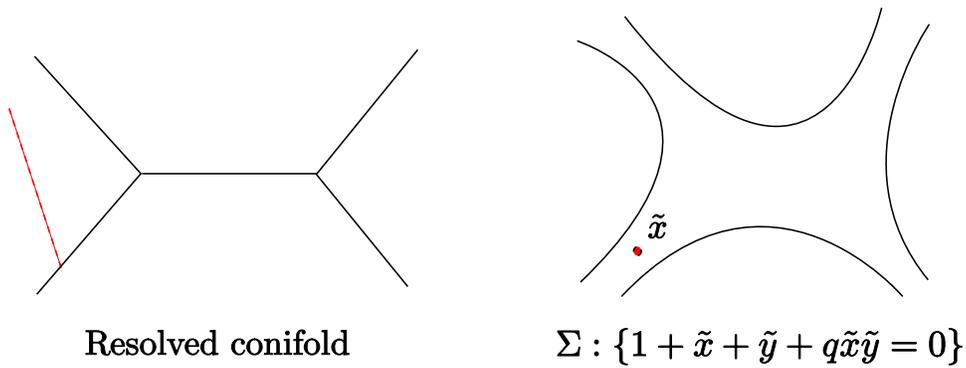}
\end{center}
\caption{The resolved conifold and its mirror curve.}
\label{f:conifold}
\end{figure}

Let us now turn to the resolved conifold, or local $\IP^1$. The mirror curve $\S \subset \IC^* \times \IC^*$ has genus $0$, and reads
\begin{equation}
H(\tx,\ty;q) = 1 + \tx + \ty + q \tx \ty,
\end{equation}
with $\tx,\ty \in \IC^*$ and $q=e^{-t}$, with $t$ the complexified K\"ahler parameter controlling the size of the $\IP^1$. This is shown in figure \ref{f:conifold}.

There are two differences with the framed vertex. First, the mirror curve above has a one-dimensional complex structure moduli space, parameterized by $q$. Hence, we could consider phase transitions in the closed moduli space. However, as explained in section \ref{moving}, since the curve has genus $0$, the amplitudes are rational functions of the closed moduli, that is there is no non-trivial monodromy for the periods. Hence, in this case these transitions are not very interesting.

Another difference is that in contrast with the framed vertex, changing phase in the open moduli space, that is, moving the brane from one leg to another, yields different amplitudes. There are basically two types of amplitudes, corresponding to ``outer" branes (ending on an outer leg of the toric diagram) and ``inner" branes, as explained in section \ref{modulispace}.  Since this type of transitions will be studied in detail for the local $\IP^2$ example, for the sake of brevity we will not present here the calculations for the resolved conifold. Let us simply mention that we checked that both the framed outer and framed inner brane amplitudes at large radius (in the limit $q \to 0$) reproduce precisely the results obtained through the topological vertex. The calculations are available upon request.

\section{Genus $1$ examples}

We now turn to the more interesting cases where the mirror curve has genus $1$. We will study two examples in detail: local $\IP^2$ and local $F_n$, $n=0,1,2$, where $F_n$ is the $n$'th Hirzebruch surface. Note that $F_0 = \IP^1 \times \IP^1$. For the sake of brevity, we do not include here all the calculations; but we are happy to provide them with more detailed explanations to the interested reader.

\subsection{Local $\IP^2$}

\begin{figure}
\begin{center}
\includegraphics[width=5in]{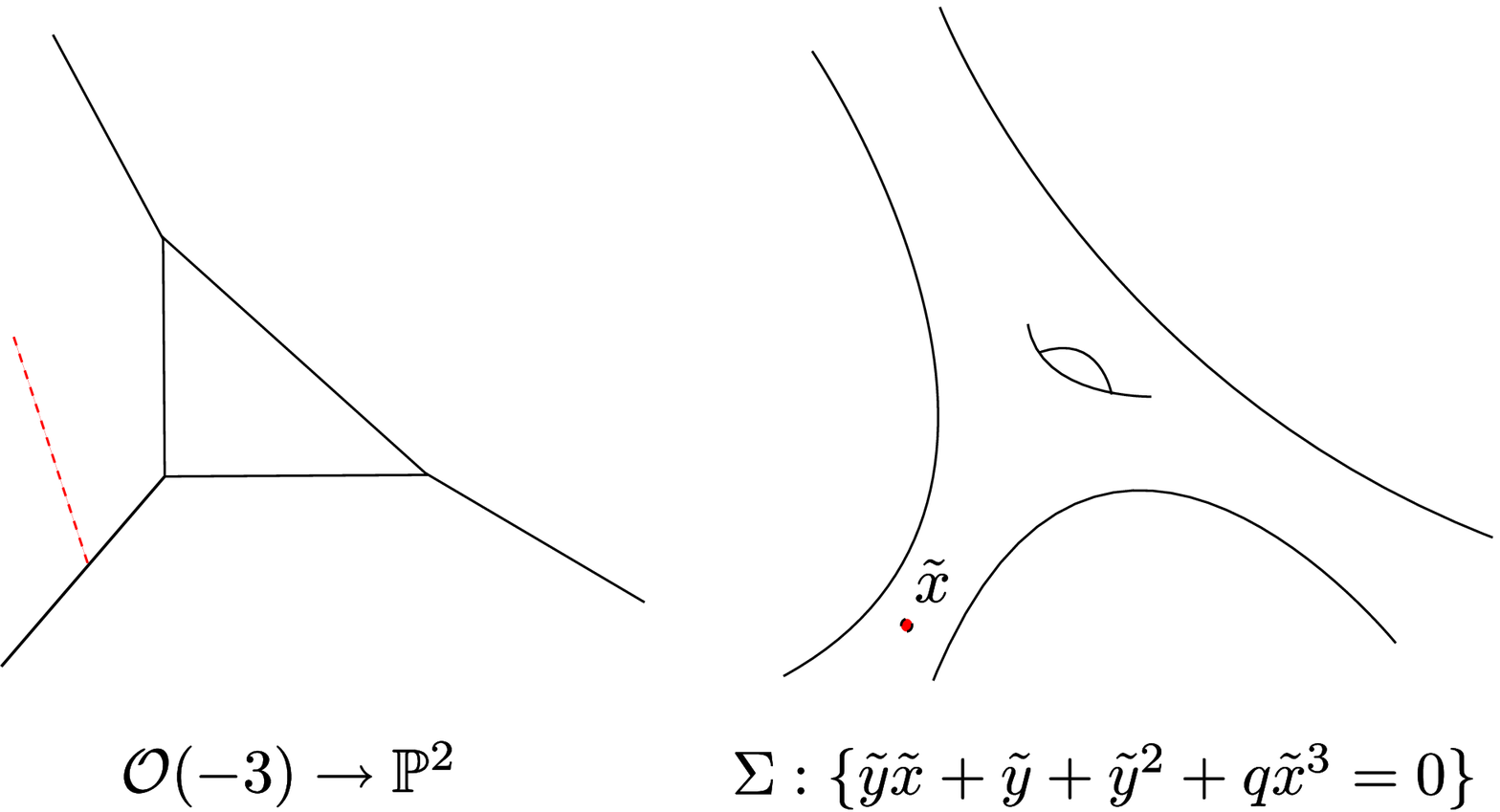}
\end{center}
\caption{An outer brane in local $\IP^2$ and its mirror.}
\label{f:localP2}
\end{figure}

The local $\IP^2$ geometry is described by the charge vector $(-3,1,1,1)$. The mirror curve is an elliptic curve with three holes, and can be written algebraically as:
\be
\label{p2in}
H(\tx_i,\ty_i;q)=\tx_i \ty_i+\tx_i^2 \ty_i+\tx_i \ty_i^2+q,
\ee
with $\tx_i, \ty_i \in \IC^*$ and $q=\re^{-t}$, with $t$ the complexified K\"ahler parameter of local $\IP^2$. As for the resolved conifold, there are two distinct phases in the open moduli space, corresponding to outer and inner branes. The above parameterization of the curve corresponds to a brane ending on an inner leg of the toric diagram, with zero framing (in standard conventions), hence the $i$ subscript. For an outer brane with zero framing, the curve reads (see section \ref{mirrormapP2})
\be
\label{p2ou}
H(\tx,\ty;q)=\ty^2+\ty+ \ty \tx + q \tx^3=0.
\ee
The outer brane geometry is shown in figure \ref{f:localP2}.

Note that as for the resolved conifold, there are now more than one phases in the closed moduli space as well. Since the curve has genus $1$, the periods now have non-trivial monodromy, and undergoing phase transitions in the closed moduli space becomes relevant. For instance, the closed moduli space contains a patch corresponding to the orbifold $\IC^3 / \IZ_3$, in the limit where the $\IP^2$ shrinks to zero size. However, in this section we will focus on the large radius limit $q \to 0$ in order to compare with the topological vertex results on the A-model side.

The mirror maps for this geometry at large radius were studied in section \ref{mirrormapP2}, for both framed outer and framed inner branes. 

\subsubsection{Framed outer amplitudes}

We start by computing the amplitudes for framed outer branes. To compute the disk amplitude we need $y = y(x)$. We get
\be
\label{invey}
y = 1 + x - f\,{x}^2 + \frac{\left( f + 3\,f^2 - 2\,z \right) \,{x}^3}{2} - 
  \frac{\left( 1 + 4\,f \right) \,\left( f + 2\,f^2 - 3\,z \right) \,{x}^4}{3} +\ldots
\ee
By definition, the bare disk amplitude is given by
\be
A_1^{(0)} (x) = \int \log y(x) {\rd x \over x},
\ee
and after expressing the result in flat open and closed coordinates using \eqref{opmap} we get precisely the topological vertex result for the disk amplitude of a framed brane in an outer leg.

We now turn to the annulus amplitude. The bare annulus amplitude is given  by:
\be
A_2^{(0)}(x_1, x_2) = \left( \int  B(x_1, x_2)\right) - \log(-x_1 + x_2).
\ee
Hence, we need the Bergmann kernel $B(x_1, x_2) $ of the framed outer curve. As explained earlier, this is simply given by reparameterizing the 
Bergmann kernel of the unframed outer curve \eqref{p2ou}.

It turns out that the unframed outer curve \eqref{p2ou} is hyperelliptic. Consequently, we can use Akemann's expression \eqref{akemann} for the Bergmann kernel in terms of the branch points of the $\tx$-projection --- here we follow the calculation performed in \cite{mm2}. To obtain these branch points, we first solve \eqref{p2ou} for $\ty$ as:
\be
\label{hycu}
\ty_{\pm}=\frac{(\tx+1)\pm \sqrt{(\tx+1)^2-4 q \tx^3} } {2}\, .
\ee
It turns out to be easier to work with the inverted variable $s = \tx^{-1}$. In this variable, the branch points of the curve are $s_1=0$ and the roots of the cubic equation
\be
s(s+1)^2-4 q=0.
\ee
In terms of
\be
\xi=\Bigl( 1 + 54\, q + 6\,{\sqrt{3}}\,{\sqrt{q\,\left( 1 + 27\,q\right) }}\Bigr)^{1\over 3},
\ee
they are given by
\be
s_2=-{2\over 3} + {1 \over 3} \Bigl(\omega  \xi+{1\over \omega \xi}\Bigr), \quad s_3=-{2\over 3} + {1 \over 3} \Bigl(\omega^* \xi +{1\over \omega^* \xi}\Bigr), \quad 
s_4={(\xi-1)^2 \over 3\xi},
\ee
where $\omega=\exp(2 \ri \pi/3)$. Plugging in these branch points in Akemann's formula \eqref{akemann}, we obtain the Bergmann kernel for the unframed outer curve, and the annulus amplitude in zero framing, as in \cite{mm2}.

We now want to implement the framing reparameterization. The reparameterization  $\tx=\tx(x)$ can be computed using that
$\tx(x)=x  {y(x)}^{-f}$ with $y(x)$ 
given in \eqref{invey}. We finally obtain, after reparameterizing the Bergmann kernel, plugging in the mirror maps \eqref{opmap}, and integrating, that the framed annulus amplitude for outer branes reads:
\begin{multline}
\label{annP2res}
\!\!\!\!\!\!\!\!\!A_2^{(0)}(X_1,X_2) = \Big[      \frac{f}{2} + \frac{f^2}{2} - \left(1 + 2\,f + 2\,f^2 \right) \,Q + \left( 4 + 7\,f + 7\,f^2 \right) \,Q^2 - \\-
  \left( 35+ 42\,f + 42\,f^2 \right) \,Q^3    + \ldots\Big] X_1 X_2 \\
+ \Big[\frac{-f}{3} - f^2 - \frac{2\,f^3}{3} + \left( 1 + 4\,f + 6\,f^2 + 4\,f^3 \right) \,Q + \left( -3 - 15\,f - 27\,f^2 - 18\,f^3 \right) \,Q^2 + \\+
  \left( 24 + \frac{308\,f}{3} + 164\,f^2 + \frac{328\,f^3}{3} \right) \,Q^3  +\cdots \Big] (X_1^2 X_2 + X_1 X_2^2)  + \ldots 
\end{multline}
This is again precisely the result obtained through the topological vertex.

The genus $0$ three-hole amplitude for framed outer branes can be computed using the general formula \eqref{3hgen}, after reparameterizing the Bergmann kernel. However, to implement this formula we need to find the branch points of the framed curve --- note that these are different from the branch points of the unframed curve found previously. 
As explained earlier, these branch points are given by the solutions of equation \eqref{bpm}.
In this case, \eqref{bpm} becomes  a cubic equation in $x$, and the three branch points  can be
determined  exactly by Cardano's method. Note that it will be relevant  which branch  of  \eqref{hycu} the branch points  belong to; thus we will use the indices $\pm$ accordingly.

The first orders of the $q$-expansion of the branch points read:\footnote{Note that the branch points
 are not regular as $f \rightarrow 0$, but the final expression of the three-hole amplitude will be.}

\begin{eqnarray}
\nonumber
&&\lambda^+_1=\frac{2 + 6\,f + 3\,f^2}{1 + 3\,f + 2\,f^2} + \frac{1 + 3\,f + 2\,f^2}{{\left( 2 + 3\,f \right) }^2\,q} - 
  \frac{{\left( 2 + 3\,f \right) }^2\,\left( 3 + 18\,f + 37\,f^2 + 30\,f^3 + 9\,f^4 \right) \,q}{{\left( 1 + 3\,f + 2\,f^2 \right) }^3} + \ldots \\     \nonumber
   &&\lambda^-_2=    \frac{1 + 3\,f}{-1 - 2\,f} + \frac{{\left( 1 + 3\,f \right) }^3\,\left( 2 + 3\,f \right) \,q}{f\,{\left( 1 + 2\,f \right) }^3} - 
  \frac{{\left( 1 + 3\,f \right) }^5\,\left( 2 + 3\,f \right) \,\left( -1 + 2\,f + 6\,f^2 \right) \,q^2}{f^3\,{\left( 1 + 2\,f \right) }^5} + \ldots   \\
 &&  \lambda_3^+=  - \frac{1}{1 + f}   + \frac{\left( -2 - 3\,f \right) \,q}{f\,{\left( 1 + f \right) }^3} - 
  \frac{ \left(2 + 3\,f \right)\,\left( 1 + f\,\left( 8 + 9\,f \right)  \right) \,q^2}{f^3\,{\left( 1 + f \right) }^5} +\ldots        \end{eqnarray}

Taking into account the branches, plugging in the mirror map and integrating, we obtain the following result in flat coordinates:
\begin{eqnarray}
&&A_3^{(0)} (X_1,X_2,X_3)=
\Big[\-\left( f^2\,{\left( 1 + f \right) }^2 \right)  + \left( 1 + 6\,f + 12\,f^2 + 12\,f^3 + 6\,f^4 \right) \,Q - \\   &&-3\,{\left( 1 + 3\,f + 3\,f^2 \right) }^2\,Q^2 + 
  4\,\left( 9 + 36\,f + 77\,f^2 + 82\,f^3 + 41\,f^4 \right) \,Q^3  +\cdots  \Big] X_1   X_2  X_3+\cdots \nonumber
 \end{eqnarray}
which reproduces again the topological vertex result.

Note that we also computed the genus $1$, one-hole amplitude, which also matches with topological vertex calculations.

\subsubsection{Framed inner amplitudes}

We can compute the amplitudes for framed inner branes in a way similar to the calculations above for outer branes. The main subtelty occurs in the reparameterization of the Bergmann kernel.

Since we want to use Akemann's formula for the Bergmann kernel, we start again with the curve in hyperelliptic form \eqref{p2ou}, which corresponds to the unframed outer brane. We then reparameterize that curve to obtain the Bergmann kernel corresponding to the curve associated to framed inner branes.

Recall that the transformation which takes the unframed outer curve to the unframed inner curve is given by \eqref{intout},
\be
(\tx, \ty) = \left( {1 \over \tx_i}, {\ty_i \over \tx_i} \right).
\ee
The framing transformation for inner branes is
\be
(\tx_i, \ty_i) = ( x_i y_i^{-f}, y_i ).
\ee
Hence we obtain the combined transformation
\be
(\tx, \ty) = (x_i^{-1} y_i^f, x_i^{-1} y_i^{f+1}),
\ee
which we can use to reparameterize the Bergmann kernel. Note that this is similar to the calculation for the framed vertex in two legs. More explicitly, we obtain
\be
\label{toinmap}
\tx(x_i)=f + \frac{1}{x_i} - \left( \frac{f}{2} + \frac{f^2}{2} \right) \, x_i + 
  \left( \frac{f}{3} + f^2 + \frac{2\,f^3}{3} \right) \,{x_i}^2 +\ldots 
\ee
Using this reparametrization and the mirror map \eqref{inmap} for framed inner branes we obtain the framed inner brane annulus amplitude: 
\begin{multline}
\!\!\!\!\!\!\!\!\!A_2^{(0)}(X_1,X_2) = \Big[     \frac{f}{2} + \frac{f^2}{2} + \left( -1 - 2\,f - 2\,f^2 - 2\,f^3 - f^4 \right) \,Q + \\ +
  \left( 11 + \frac{35\,f}{2} + \frac{81\,f^2}{4} + \frac{157\,f^3}{8} + \frac{93\,f^4}{8} + \frac{27\,f^5}{8} + 
     \frac{5\,f^6}{8} \right) \,Q^2  - \Big( 131 +201\,f +\\+ \frac{467\,f^2}{2} + \frac{15023\,f^3}{90} + 
     \frac{781\,f^4}{180} - \frac{47\,f^5}{72} + \frac{1429\,f^6}{18} - \frac{1537\,f^7}{360} - \frac{221\,f^8}{180} \Big) \,
   Q^3   + \ldots\Big] X_1 X_2 \\ + \Big[\left( 1 - \frac{3\,f}{2} + \frac{f^2}{2} \right) \,Q^2 + \left( -8 + 16\,f - 14\,f^2 + 6\,f^3 - f^4 \right) \,Q^3+\ldots \Big] \frac{1}{X_1 X_2}+ \ldots 
\end{multline}
This reproduces the topological vertex result, including both positive and negative winding numbers contributions.

Note that we also computed the genus $0$, three-hole and the genus $1$, one-hole amplitudes for framed inner amplitudes and obtained perfect match again. 

We also computed the annulus amplitude for one brane in an outer leg and one brane in an inner leg, paralleling the framed vertex in two legs calculation. We again obtained perfect agreement.

\subsection{Local $F_n$, $n=0,1,2$}

\label{fn}

We now study the local $F_n$, $n=0,1,2$ geometries, where $F_n$ is the $n$'th Hirzebruch surface. Note that $F_0 = \IP^1 \times \IP^1$.

\begin{figure}
\begin{center}
\includegraphics[width=5in]{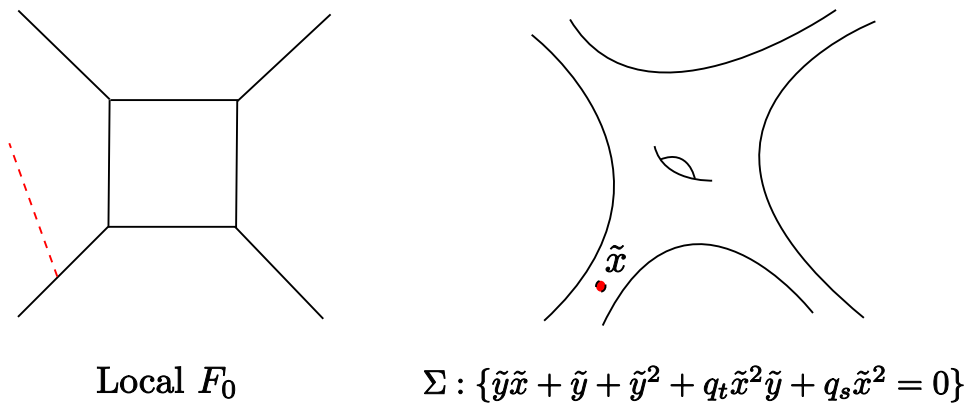}
\end{center}
\caption{An outer brane in local $F_0$ and its mirror.}
\label{f:hirzebruch}
\end{figure}

The local $F_n$ geometries are described by the two charge vectors:
\begin{align}
Q^1 =&( -2, 1, 1, 0, 0),\notag\\
Q^2 =&(n-2, 0,-n, 1, 1).
\end{align}
The mirror curves $\S_n \subset \IC^* \times \IC^*$ have genus $1$ and four punctures. In the paramaterization corresponding to a brane placed in an external leg (with zero framing), they  read:
\be
\label{fnou}
H_n(\tx,\ty;q_t,q_s)=\ty \tx + \ty + \ty^2 + q_t  
\tx^2 \ty + q_t^n  q_s \tx^{n+2}\, ,
\ee
with $\tx,\ty \in \IC^*$,  $q_t=e^{-t}$ and $q_s=e^{-s}$, with $t$ and $s$ the complexified K\"ahler parameters. The local $F_0$ geometry is shown in figure \ref{f:hirzebruch}.

The closed moduli space is now two-dimensional, being spanned by $q_t$ and $q_s$. However, these curves are still hyperelliptic, 
and we can apply our formalism exactly as we did for the local  $\IP^2$ case. Therefore, we will not do the full calculation here, but only highlight some interesting aspects.

The large radius expansion for local $F_0=\IP^1 \times \IP^1$
has been discussed in detail in \cite{mm2},  where several open amplitudes 
(for  outer  branes with  canonical framing)  were computed. Needless to say,
we checked that our formalism can be used to complete the calculations by including framing
and inner brane configurations.

Besides the large radius point, our formalism
allows to compute topological strings amplitudes at other points in
 the closed moduli space of local $F_0$,  like the  conifold point and the  orbifold point. The latter  corresponds  to the point  where the $\IP^1 \times \IP^1$ shrinks to zero size. This special point will be discussed in great detail in the next section.

For local $F_1$ and $F_2$, the open and closed mirror maps together 
with the disk amplitudes for inner and outer branes were studied, for instance, in \cite{LM}.
Again, we showed that our formalism allows to compute framed inner and outer higher amplitudes
in the large radius limit, checking  our  results with the topological vertex ones.

As an example, the outer annulus amplitude at zero framing for the  local $F_1$ geometry reads:
\be
\ba
\label{annf1}
&A_2^{(0)}(X_1, X_2) =\Bigl[-Q_s Q_t-3 Q_t^2 Q_s  + 4  Q_t^2 Q_s^2 - 5  Q_s Q_t^3+\cdots\Bigr] X_1 X_2 \\
&-\Bigl[ -Q_s Q_t-2 Q_t^2 Q_s  + 3  Q_t^2 Q_s^2 - 4  Q_s Q_t^3+\cdots \Bigr]\left(X_1 X_2^2+X_1^2 X_2\right)  \\
&+\Bigl[Q_s Q_t-2 Q_t^2 Q_s  + 4  Q_t^2 Q_s^2 - 3  Q_s Q_t^3+\cdots\Bigr] \left(X_1 X_2^3+X_1^3 X_2\right)  \\
&+\Bigl[-Q_s Q_t-2 Q_t^2 Q_s  + \frac{7}{2}  Q_t^2 Q_s^2 - 3  Q_s Q_t^3+\cdots \Bigr] X_1^2 X_2^2   +\cdots
\ea
\ee
while for the local  $F_2$  geometry:
\be
\ba
&A_2^{(0)}(X_1, X_2) =\Bigl[  2 Q_t^2 Q_s  + 4  Q_t^3 Q_s + \cdots\Bigr] X_1 X_2 
-\Bigl[ Q_t^2 Q_s +3   Q_t^3 Q_s +\cdots \Bigr] \left(X_1 X_2^2+X_1^2 X_2\right)  \\
&+\Bigl[ Q_t^2 Q_s + 2Q_t^3 Q_s +\cdots\Bigr] \left(X_1 X_2^3+X_1^3 X_2\right)
+\Bigl[ Q_t^2  Q_s + 2 Q_s Q_t^3 +\cdots \Bigr] X_1^2 X_2^2   +\cdots
\ea
\label{annpone}
\ee
Both of these coincide indeed with the topological vertex results.

There is also another interesting phase in the local $F_1$ moduli space --- see for instance \cite{IKP}. By definition, $F_1$ is a $\IP^1$ bundle over $\IP^1$, where the $\IP^1$ base is an exceptional curve. In fact, $F_1$ is isomorphic to $\IP^2$ blown up in one point, the base of the fibration corresponding to the blown up exceptional curve. Hence, we can blow down this exceptional $\IP^1$, and we should recover $\IP^2$. In other words, if we take the open amplitudes for local $F_1$ and move to the phase in the moduli space where this exceptional $\IP^1$ goes to zero size, we should recover the open amplitudes for local $\IP^2$. Going to this patch in fact corresponds to a mild transformation in the closed moduli space, since it does not involve a redefinition of the periods. The phase transition can then be directly implemented on the
amplitudes as no  modular transformation is needed.

More specifically, it can be implemented in the local $F_1$ annulus amplitude \eqref{annf1} by first defining $\tilde Q_s=Q_s Q_t $ and then taking the limit $Q_t \rightarrow 0$. We get:
\be
\ba
&A_2^{(0)}(X_1, X_2) =\Bigl[  -\tilde Q_s+ 4  \tilde Q_s^2  + \cdots\Bigr] X_1 X_2 
+\Bigl[ \tilde Q_s   - 3  \tilde Q_s^2 +\cdots \Bigr]\left(X_1 X_2^2+X_1^2 X_2\right)  \\
&+\Bigl[ -\tilde Q_s   + 4  \tilde Q_s^2  +\cdots\Bigr] \left(X_1 X_2^3+X_1^3 X_2\right)  
+\Bigl[ -\tilde Q_s   + \frac{7}{2}  \tilde Q_s^2  +\cdots \Bigr]X_1^2 X_2^2   +\cdots
\ea
\ee
which indeed coincides with the local  $\IP^2$ annulus amplitude at zero framing, see \eqref{annP2res}.

\section{Orbifold points}

\label{orbifold}

As we already emphasized, one of the main feature of our B-model formalism is that it can be used to study various phases in the open/closed moduli space, not just large radius points. In particular, there are two special points where we can use our formalism to generate open and closed amplitudes; the orbifold point of local $\IP^2$, which corresponds to the orbifold $\IC^3 / \IZ_3$, and the point in the moduli space of local $\IP^1 \times \IP^1$ where the $\IP^1 \times \IP^1$ shrinks to zero size (which we will call the local $\IP^1 \times \IP^1$ orbifold point, although it is not really an orbifold). 

In the second example, we can use large $N$ dualities to make a precise test of our formalism, and of its ability to 
produce results in all of the K\"ahler moduli space (and not only at the large radius limit). Indeed, it was argued in \cite{akmvmm} that topological strings on $A_{p-1}$ fibrations 
over $\IP^1$ are dual to Chern--Simons theory on the lens space 
$L(p,1)$. In particular, the topological string 
expansion around the orbifold point of these geometries can be computed by doing perturbation theory in the 
Chern--Simons gauge theory. This was checked for closed string amplitudes in \cite{akmvmm}, for $p=2$. We will extend 
this duality to the open string sector and make a detailed comparison of the amplitudes. 

In the first example, we would obtain open and closed orbifold amplitudes of $\IC^3 / \IZ_3$. The closed amplitudes were already studied in \cite{abk}; by now some of the predictions of that paper for closed orbifold Gromov-Witten invariants have been proved mathematically. For the open amplitudes, to the best of our knowledge open orbifold Gromov-Witten invariants have not been defined mathematically, hence there is nothing to compare to. However, we proposed in section \ref{phaseflat} a method for determining the flat coordinates at all degeneration points in the moduli space which, as we will see, applies to the local $\IP^1 \times \IP^1$ orbifold point. Therefore, we will assume that it should work at the $\IC^3 / \IZ_3$ orbifold point as well, and use it to make predictions for the disk amplitude for $\IC^3 / \IZ_3$.

Let us start by studying the local $\IP^1 \times \IP^1$ orbifold point.  We first perform the Chern-Simons calculation, then explain the large $N$ duality, and finally present our dual B-model calculation.

\subsection{Chern--Simons theory and knots in lens spaces}

In order to extend the duality of \cite{akmvmm} to the open sector, we will need some detailed computations 
in Chern--Simons theory. In this subsection we review \cite{mm,akmvmm} and extend them slightly to include Wilson loops. 

Lens spaces of the form $L(p,1)$ can be obtained by gluing two solid 2-tori along their 
boundaries after performing the 
${\rm SL}(2,\IZ)$ transformation,
\be
\label{lensmat}
U_p =\begin{pmatrix}1 &  0 \\
p & 1 \end{pmatrix}.
\ee
This surgery description makes it possible to calculate the partition function of Chern--Simons theory on these 
spaces, as well as correlation functions of Wilson lines along trivial knots, in a simple way. To see this, we first recall some 
elementary facts about Chern--Simons theory. 
  
An ${\rm SL}(2,\IZ)$ transformation given by the matrix
\begin{equation}
\label{sltwo}
U^{(p_i,q_i)}=\left( \begin{array} {cc} p_i & r_i \\ 
q_i & s_i\end{array} \right)
\end{equation}
lifts to an operator acting on $\CH(\IT^2)$, the Hilbert space obtained by 
canonical quantization of Chern--Simons theory on the 2-torus. This space is the 
space of integrable representations of a WZW model with gauge group $G$ at
level $k$, where $G$ and $k$ are respectively the Chern--Simons gauge group and the 
quantized coupling constant. We will use the following notations: 
$r$ denotes 
the rank of $G$, and $d$ its dimension. 
$y$ denotes the dual Coxeter number. 
The fundamental weights will be denoted by
$\lambda_i$, and the simple roots by $\alpha_i$, with $i=1, \cdots, r$. 
The weight and root lattices of $G$ are
denoted by $\Lambda_{\rm w}$ and $\Lambda_{\rm r}$, respectively. 
Finally, we put $l=k +y$. 

Recall that a representation given by a highest weight $\Lambda$ is integrable if 
the weight $\rho + \Lambda$ is in the   
fundamental chamber ${\cal F}_l$ ($\rho$ denotes as usual the Weyl vector,
given by the sum of the fundamental weights). The fundamental chamber 
is given by $\Lambda_{\rm w}/l 
\Lambda_{\rm r}$ modded out by the action of the Weyl group. For example,
in $SU(N)$ a 
weight $p=\sum_{i=1}^r p_i \lambda_i$ is in ${\cal F}_l$ if  
\begin{equation}
\sum_{i=1}^r p_i < l,\,\,\,\,\,\, {\rm and} \,\,\ p_i >0, \, i=1, \cdots, r.
\end{equation}
In the following, 
the basis of integrable representations will be labeled by the 
weights in ${\cal F}_l$.  

In the case of simply-laced gauge groups, 
the ${\rm SL}(2, \IZ)$ transformation given by $U^{(p,q)}$ has
the following matrix elements in the above basis \cite{roz,ht}:
\be
\ba
&\langle \alpha |\CU^{(p, q)}| \beta\rangle = 
{[i \, {\rm sign}(q)]^{|\Delta_+|} \over (l |q|)^{r/2}}
\exp \Bigl[ -{ i d \pi \over 12}  \Phi (U^{(p,q)})\Bigr] 
\Biggl( { {\rm Vol}\, \Lambda_{\rm w} \over {\rm Vol}\,  \Lambda_{\rm r}} 
\Biggr)^{1 \over 2} \\
& \cdot \sum_{n \in \Lambda_{\rm r}/q \Lambda_{\rm r}}\sum_{w \in {\cal W}} \epsilon
(w) \exp \Bigl\{ {i \pi \over l q} \Bigl( p \alpha^2 - 2\alpha (l n + w(\beta))
+ s(ln + w(\beta))^2 \Bigr)\Bigr\}.
\ea
\ee
In this equation, $|\Delta_+|$ denotes the number of positive roots of $G$, 
and the second sum is over the Weyl group ${\cal W}$ of $G$.
 $\Phi (U^{(p,q)})$ is the Rademacher function:
\begin{equation}
\label{rade}
\Phi\left[ \begin{array}{cc} p & r \\ q&s\end{array} \right]= 
{p + s \over q} - 12 s(p,q),
\end{equation}
where $s(p,q)$ is the Dedekind sum
\begin{equation}
s(p,q)={1 \over 4q} \sum_{n=1}^{q-1} \cot \Bigl( {\pi n \over q}\Bigr) 
\cot \Bigl( {\pi n p\over q}\Bigr).
\end{equation} 

From the above description it follows that the partition function of the lens space $L(p,1)$ is given by
\be
Z(L(p,1)) = \langle \rho |\CU_p |\rho\rangle, 
\ee
where $\CU_p$ is the lift of (\ref{lensmat}) to an operator on $\CH(\IT^2)$. In order to make contact with the open sector, we 
need as well the normalized vacuum expectation value of a Wilson line along the unknot in $L(p,1)$, in the 
representation $R$, which is given by
\be
W_R={\langle \rho| \CU_p | \rho +\Lambda\rangle \over \langle \rho |\CU_p |\rho\rangle}, 
\ee
where $\Lambda$ is the highest weight corresponding to $R$. The numerator can be written (up to an overall constant that will cancel with the denominator)
\be
 \sum_{n \in \Lambda_{\rm r}/p \Lambda_{\rm r}}\sum_{w \in {\cal W}} \epsilon
(w) \exp \Bigl\{ {i \pi \over l p} \Bigl( \rho^2 - 2\rho (l n + w(\rho+\Lambda))
+ (ln + w(\rho+\Lambda))^2 \Bigr) \Bigr\}.
\ee
It is a simple exercise in Gaussian integration to check that this quantity can be written as
\be
\sum_{n \in \Lambda_{\rm r}/p \Lambda_{\rm r}} \sum_{w,w'\in \CW} \epsilon(w w') \int \prod_{i=1}^r \rd \lambda_i \exp\Bigl\{ 
-{1\over 2 \hat g_s}  \lambda^2 -\ell n \cdot \lambda+ \lambda \cdot (w(\rho)-w'(\rho +\Lambda))\Bigr\} ,
\ee
where 
\be
\label{cscoup}
\hat g_s={2\pi \ri \over p l},
\ee
and $\rd \lambda= \prod_{i=1}^r \rd \lambda_i$ and $\lambda_i$ are the Dynkin coordinates of $\lambda$, understood as an element in $\Lambda_w\otimes \IR$. 
This integral can be further written as
\be
\int \rd \lambda \exp \Bigl \{ -{1\over 2 \hat g_s}  \lambda^2 -\ell n \cdot \lambda\Bigr\} \prod_{\alpha>0} \biggl( 2 \sinh {\lambda \cdot \alpha \over 2}\biggr)^2 \tr_R \, \re^{-\lambda}, 
\ee
where we have used Weyl's formula for the character, 
\be
\tr_R \, \re^{-\lambda}={\sum_{w\in \CW} \epsilon(w) \re^{-\lambda \cdot w(\rho+\Lambda)} \over \sum_{w\in \CW} \epsilon(w) \re^{-\lambda \cdot w(\rho)} }
\ee
as well as Weyl's denominator formula. It follows that 
\be
\label{wrintf}
W_R={1\over Z(L(p,1))} \int \rd \lambda \exp \Bigl \{ -{1\over 2 \hat g_s}  \lambda^2 -\ell n \cdot \lambda\Bigr\} \prod_{\alpha>0} \biggl( 2 \sinh {\lambda \cdot \alpha \over 2}\biggr)^2 \tr_R \, \re^{-\lambda}, 
\ee
where 
\be
\label{pfint}
Z(L(p,1))=\int \rd \lambda \exp \Bigl \{ -{1\over 2 \hat g_s}  \lambda^2 -\ell n \cdot \lambda\Bigr\} \prod_{\alpha>0} \biggl( 2 \sinh {\lambda \cdot \alpha \over 2}\biggr)^2.
\ee
This provides matrix integral representations for both the partition function (derived previously in \cite{mm,akmvmm}) and the normalized vacuum expectation value of 
a Wilson line around the unknot. Both expressions are computed in the background of an arbitrary flat connection labelled by the vector $n$. Notice that, when $n=0$, 
one has that
\be
W_R=\re^{\hat g_s/2(\kappa_R +\ell(R)N)} {\rm dim}_q \, R,
\ee
where ${\rm dim}_q \, R$ is the $U(N)$ quantum dimension of $R$ with $q=\re^{\hat g_s}$. We can therefore regard (\ref{wrintf}) 
for arbitrary $n$ as a generalization of quantum dimensions. 

As shown in \cite{mm,akmvmm}, the partition function above can be written more conveniently in terms of a multi--matrix model 
for $p$ Hermitian matrices. In the case of $L(2,1)$ (to which we will restrict ourselves), a generic flat connection can be specified by a breaking 
$U(N) \rightarrow U(N_1) \times U(N_2)$, or equivalently by a vector $n$ with 
$N_1$ $+1$ entries and $N_2$ $-1$ entries. It is then easy to see \cite{akmvmm} that the partition function (\ref{pfint}) is 
given by the Hermitian two-matrix model, 
\be
\label{pftwoone}
Z(N_1, N_2, \hat g_s)=\int \rd M_1 \rd M_2 \exp \biggl\{ -{1 \over 2 \hat g_s} {\rm
Tr}M_1^2 -{1 \over 2 \hat g_s} {\rm
Tr}M_2^2+ V(M_1)+ V(M_2) + W(M_1, M_2)\biggr\},
\ee
where
\be
\ba
V(M)& =
{1 \over 2}\sum_{k=1}^{\infty} a_k \sum_{s=0}^{2k}
(-1)^s {2k \choose
s} {\rm Tr}M^s {\rm Tr}M^{2k-s},\\
W(M_1, M_2) &= \sum_{k=1}^{\infty} b_k \sum_{s=0}^{2k}
(-1)^s {2k \choose
s} {\rm Tr}M_1^s {\rm Tr}M_2^{2k-s},
\ea
\ee
and
\be
a_k={ B_{2k} \over k (2k)!}, \qquad b_k= {2^{2k}-1 \over k (2k)!}B_{2k}.
\ee

The vacuum expectation value of the unknot in $L(2,1)$ is similarly given by 
\be
\label{wrint}
W_{R}(N_1, N_2, \hat g_s)=  {1\over Z (N_1, N_2, \hat g_s)} \langle  \tr_{R}  \re^{M} \rangle,
\ee
where, in terms of the eigenvalues $m_i^1$, $m_j^2$ of $M_1$, $M_2$, the matrix $\re^M$ is given by 
\be
\re^M={\rm diag}(\re^{m^1_1}, \cdots, \re^{m^1_{N_1}}, -\re^{m^2_1},\cdots, -\re^{m^2_{N_2}}).
\ee
The vev in (\ref{wrint}) is defined by the weight given by the exponent in (\ref{pftwoone}), and it can be easily computed in 
perturbation theory. 

In order to compare the results with the string theory results, we will need to compute the connected vevs $W^{(c)}_{\vec k}$ 
in the $\vec k=(k_1, k_2, \cdots)$ basis, which are defined by 
\be
\log \Bigl[ \sum_R W_R {\rm Tr}_R \, V \Bigr]=\sum_{\vec k}{1 \over z_{\vec k}!}
W^{(c)}_{\vec k} \Upsilon_{\vec k}(V) ,
\end{equation}
where the notations are as in (\ref{convev}). Using the matrix model representation we can easily compute, for example, 
\be
\ba
W^{(c)}_{(1,0,\cdots)}&=N_1-N_2 + {\hat g_s\over 2} (N_1^2-N_2^2) + {\hat g_s^2 \over 24} (N_1-N_2) \Bigl(4 N_1^2 + 4 N_2^2 +10 N_1 N_2 -1\Bigr) + \cdots\\
W_{(2,0,\cdots)}^{(c)}&=\hat g_s (N_1+ N_2) + {\hat g_s^2 \over 2} \Bigl( 3 N_1^2 + 3 N_2^2 + 4 N_1 N_2 \Bigr) \\ &+ {\hat g_s^3\over 6}\Bigl( 7 (N_1^3 + N_2^3) + 15 (N_1^2 N_2+ N_1 N_2^2) \Bigr) \\
& + {\hat g_s^4 \over 24} \Bigl(15 N_1^4 + 47 N_1^3 N_2 +\{ N_1 \leftrightarrow N_2\}+ 63 N_1^2 N_2^2 + 3 N_1 N_2 \Bigr)+ \cdots \\
W^{(c)}_{(0,1,0,\cdots)}&= N_1+N_2 + 2\hat g_s (N_1^2 + N_2^2) +  {\hat g_s^2 \over 3} (N_1+ N_2)\Bigl( 5 N_1^2 + 5N_2^2 -2 N_1 N_2 +1\Bigr)\\
& +{\hat g_s^3 \over 12}\Bigl( 11 N_1^4+ 18 N_1^3 N_2  + 6 N_1^2 N_2^2 + 5 N_1^2 + 18 N_1 N_2 + 
\{ N_1 \leftrightarrow N_2\}\Bigr).
\ea
\ee
In order to compare with topological string amplitudes it is convenient to reorganize the connected vevs in terms of the 
't Hooft expansion. To do that, we introduce the 
't Hooft variables 
\be
\label{tpar}
S_i=\hat g_s N_i, \quad i=1, \cdots, p.
\ee
A diagrammatic argument based on fatgraphs says that the connected vevs have the structure
\be
W^{(c)}_{\vec k}(S_i, \hat g_s)=\sum_{g,h_i} \hat g_s^{2g-2+|\vec k| + \sum_i h_i } F_{g, \vec k, h_i} N_1^{h_1} \cdots N_p^{h_p} 
=\sum_{g,h_i} \hat g_s^{2g-2+|\vec k|} F_{g, \vec k, h_i} S_1^{h_1} \cdots S_p^{h_p}.
\ee
The explanation for this is simple: in terms of fatgraphs, the connected vev $W^{(c)}_{\vec k}(N_i,\hat g_s)$ is obtained by summing over 
fatgraphs with a fixed number of holes $|\vec k|$ but with varying genus $g$ and number of ``coloured" holes $h_i$. We can sum over 
all coloured holes at fixed genus to obtain the amplitude
\be
W^{(g)}_{\vec k}(S_i)=\sum_{h_i} F_{g, \vec k, h_i} S_1^{h_1} \cdots S_p^{h_p}. 
\ee
Finally, in order to make contact with the open toplogical string amplitudes we notice that
\be
\label{gaugevevs}
\sum_{\vec k, \, |\vec k|=h} {1\over z_{\vec k}} W^{(g)}_{\vec k}(S_i) \Upsilon_{\vec k}(V) =W_h^{(g)}(z_1, \cdots, z_h),
\ee
under the dictionary (\ref{idenwords}). 

From the above explicit computations we get the following results:
\be
\ba
\label{orbidisk}
A^{(0)}_1(p)&= p\biggl\{ S_1-S_2 + {1\over 2} (S_1^2 -S_2^2) + {1\over 24} (S_1-S_2)(4 S_1^2 +10 S_1 S_2+ 4 S_2^2) \\& + {1\over 24} (S_1-S_2) (S_1^3  + S_2^3 + 4 (S_1^2 S_2 + S_1 S_2^2)) + \cdots\biggr\} \\
&+ {1\over 2} p^2\biggl\{ S_1+ S_2 + 2 (S_1^2+ S_2^2)  +{1\over 3} (S_1+S_2)(5 S_1^2 -2 S_1 S_2+ 5 S_2^2) \\ 
&+{1\over 12} \Bigl( 11(S_1^4 + S_2^4) + 18 (S_1^3 S_2 + S_2^3 S_1) +6 S_1^2 S_2^2 \Bigr)\\
&+ {1\over 180} (S_1+ S_2)\Bigl( 69 S_1^4 + 126 S_1^3 S_2 -6 S_1^2 S_2^2 + \{ S_1 \leftrightarrow S_2\} \Bigr) + \cdots \biggr\} \\
&+ {1\over 3} p^3 \biggl\{ S_1- S_2 + {9\over 2}  (S_1^2- S_2^2)  +{1\over 4} (S_1-S_2)(30 (S_1^2 +S_2^2)  + 39 S_1 S_2) \cdots \biggr\},
\ea
\ee
\be
\ba
A^{(1)}_1(p)&=p\biggl\{ -{1\over 24}(S_1-S_2) -{1\over 48}(S^2_1-S^2_2) -{1\over 576} (S_1-S_2) \Bigl(4 (S_1^2 + S_2^2 ) + 19 S_1 S_2 \Bigr) + \cdots\biggr\} \\ 
& + {1\over 2} p^2 \biggl\{ {1\over 3}(S_1 + S_2) + {1\over 12} \Bigl( 5(S_1^2 + S_2^2)  + 18 S_1 S_2 \Bigr) + {1\over 180} (S_1 + S_2) \Bigl(55(S_1^2 + S_2^2) + 230 S_1 S_2\Bigr) \\ &\qquad \,\,\,\,\,\,\,\,\,\, +\cdots \biggr\} +{p^3 \over 3} \biggl\{ {21 \over 8}(S_1-S_2) + \cdots\biggr\} ,
\ea
\ee
\be
\ba
\label{orbian}
A^{(0)}_2(p,q)&=p q \biggl\{  S_1 + S_2 + {1\over 2} (3 S_1^2 + 3S_2^2 + 4 S_1 S_2) +{1\over 6} (S_1 + S_2) (7 (S_1^2 + S_2^2) + 8 S_1 S_2) \\
& + {1\over 24} \Bigl( 15 (S_1^4 + S_2^4) + 47 (S_1^3 S_2 + S_2^3 S_1) + 63 S_1^2 S_2^2\Bigr)+\cdots \biggr\} \\ &+
(p^2 q + pq^2) \biggl\{ (S_1 -S_2) +{7\over 2} (S_1^2 - S_2^2)  + {1\over 12}\Bigl( 62 (S_1^3-S_2^3) + 51(S_1^2 S_2 - S_1 S_2^2) \Bigr) \\ & + {1\over 24}\Bigl( 115 (S_1^4  -S_2^4) + 201 (S_1^3 S_2-S_1 S_2^3) \Bigr)+ \cdots \biggr\},
\ea
\ee
\be
A^{(0}_3 (p,q,r)=pqr\Bigl\{ 3(S_1-S_2) + {17\over 2} (S_1^2 -S_2^2) + {1\over 4} (46 (S_1^3-S_2^3) + 45 (S_1^2 S_2-S_2^2 S_1)) +\cdots \Bigr\}.
\ee

Finally, as explained in \cite{wittencs}, Wilson loop operators in Chern--Simons theory need a choice of framing in order to 
be properly defined. The calculations above correspond to the framing coming naturally from the Gaussian integral in (\ref{wrint}), and 
to change the framing by $f$ units it is enough to multiply $W_R$ by 
\be
\exp\{ -f \hat g_s \kappa_R/2\}, 
\ee
The amplitudes computed above would change correspondingly. We would have, for example, 
\be
\ba
\label{orbifrdisk}
A_1^{(0)}(p)&= p\biggl\{ S_1-S_2 +{1\over 2} (S_1^2 -S_2^2) 
+ {1\over 24} (S_1-S_2)\Bigl( 4 (S_1^2 +S_2^2)  + 10 S_1 S_2 \Bigr)+ \cdots\biggr\} \\
&+ {p^2 \over 2} \biggl\{ S_1+ S_2 +(2-f) (S_1^2+ S_2^2)  +2 f S_1 S_2 \\
&+{1\over 3} (S_1+S_2)\Bigl((5-3 f) (S_1^2  +S_2^2)+2(3f-1) S_1 S_2 \Bigr)+ \cdots \biggr\},
\ea
\ee
and
\be
A^{(0)}_2(p,q)= p q \biggl\{ (1-f) (S_1 + S_2) + {1\over 2} \Bigl( (3 -4f+f^2) (S_1^2 + S_2^2) + (4 -2 f^2) S_1 S_2\Bigr) +\cdots \biggr\} + \cdots
\ee

\subsection{The orbifold point and a large $N$ duality}

In \cite{akmvmm} it was argued that topological string theory on $X_p$, the symmetric $A_{p-1}$ fibration over $\IP^1$, is dual 
to Chern--Simons theory in the lens space $L(p,1)$. This is a highly nontrivial example of a gauge theory/string theory duality 
which can be obtained by a $\IZ_p$ orbifold of the large $N$ duality of Gopakumar and Vafa \cite{gv}. Equivalently, it can be understood 
as a geometric transition between $T^*(\IS^3/\IZ_p)$ (which is equivalent to Chern--Simons theory on $L(p,1)$ \cite{csosft}) and the $X_p$ 
geometry. 

Checking this duality is complicated because the perturbative regime of the gauge theory, where one can 
do computations easily, corresponds to string theory on $X_p$ near the point $t_i = 0$, where the $t_i$ are the K\"ahler parameters. This is 
a highly stringy phase --- a small radius region --- where the $\alpha'$ corrections are very important. It is conventional to 
refer to this point as an orbifold point (although the periods are still logarithmic) and we will do so in the following. 
This type of problem in testing the duality is well-known in the context of 
the AdS/CFT correspondence, where the perturbative regime of $\CN=4$ Yang--Mills 
corresponds to a highly curved AdS$_5\times \IS^5$ target. In order to 
proceed, one has to either do computations in the strong 't Hooft coupling regime of Chern--Simons theory, or to solve topological 
string theory near the orbifold point. Thanks to mirror symmetry and the B-model, the second option is  easier, and 
this was the strategy used in \cite{akmvmm} to test the duality in the closed string sector.

\begin{figure}
\leavevmode
 \begin{center}
\epsfxsize=5in
\epsfysize=2in
\epsfbox{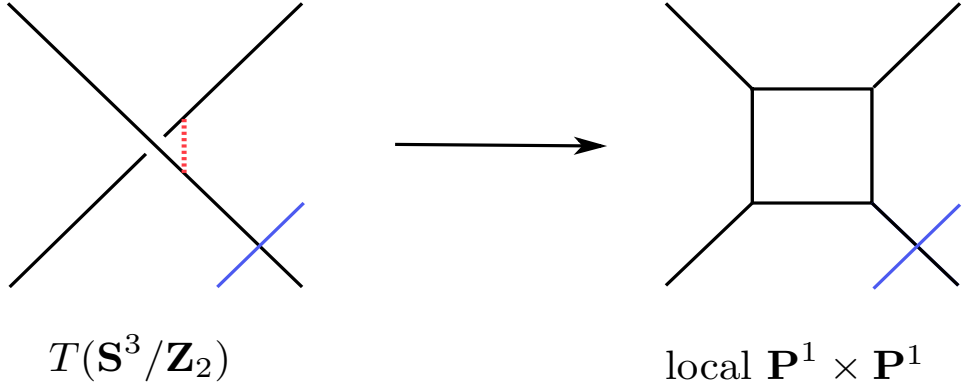}
\end{center}
\caption{The geometric transition between $T^*(\IS_3/\IZ_2)$ with a Lagrangian brane associated to the unknot, and 
local $\IP^1 \times \IP^1$ with an outer brane.}
\label{transition}
\end{figure}

How would we extend this story to the open sector? First we recall that, in the Gopakumar--Vafa duality, 
a knot $\CK$ in $\IS^3$ leading to a Wilson loop operator in Chern--Simons gauge theory corresponds 
to a Lagrangian submanifold $\CL_{\CK}$ in the resolved conifold \cite{ov}. Moreover, the connected vevs (\ref{gaugevevs}) become, under this duality,   
open string amplitudes with the boundary conditions set by $\CL_{\CK}$. After orbifolding by $\IZ_p$, the natural statement (generalizing the 
results of Ooguri and Vafa in \cite{ov}) is that a knot in $L(p,1)$ corresponds to a Lagrangian submanifold in $X_p$. The simplest test of the 
Ooguri--Vafa conjecture is the unknot, which corresponds to a toric D-brane in an outer edge of the resolved conifold (see for example \cite{mmrev} for 
details). It is then natural to conjecture that {\it the unknot in $L(p,1)$ is dual to a toric D-brane in an outer edge of $X_p$}, and that the connected vevs 
for the corresponding Wilson line correspond to open string amplitudes for this brane. This should follow from the geometric transition for $X_p$ proposed in 
\cite{akmvmm}, and it is sketched in \figref{transition}.

Testing this conjecture is again difficult for the reasons explained above. In order to compare with the perturbative string amplitudes that we computed from the 
Chern--Simons matrix model, we need a way to compute open string amplitudes that makes it possible to go anywhere in the moduli space. But this is precisely one of the 
outcomes of the B-model formalism proposed in this paper! We will now explain how to compute open string amplitudes in the $p=2$ case, {\it i.e.} local $\IP^1 \times \IP^1$, near 
the orbifold point, extending in this way the test of the duality performed in \cite{akmvmm} to the open sector. This will verify not only our extension of the duality for knots 
in the lens space $L(2,1)$, but also the power of our B-model formalism. 


\subsection{Orbifold  Amplitudes}

We now explain how to compute  open string amplitudes at the
orbifold point  in the local $\IP^1 \times \IP^1$ geometry, using the B-model formalism developed in this paper. We follow the general discussion in section \ref{movingBmodel}. 
Basically, to compute the open amplitudes at the orbifold point, one only needs to find the disk and the annulus amplitudes at this point, and then use our B-model formalism to generate the other amplitudes recursively. We also need to fix the open and closed mirror maps at the orbifold point in order to compare with the Chern-Simons results.

Let us start by introducing the geometrical data, as in section \ref{fn}. The two charge vectors for local $\IP^1 \times \IP^1$ are:
\begin{equation}
\begin{array}{rl}
Q^1&=(-2,1,1,0,0),\\
Q^2&=(-2,0,0,1,1) .
\end{array}
\end{equation}
The mirror curve in the parameterization corresponding to an
outer brane with zero framing is hyperelliptic and reads:
\begin{equation}
\label{p1p1} H(\tx,\ty;q_s,q_t)=\ty^2+\ty\,(q_s \tx^2+1+\tx)+ q_t \tx^2 \, ,
\end{equation}
with $q_s = \re^{-t_s}$ and $q_t = \re^{-t_t}$.

This geometry was studied at large radius in \cite{mm2}. Solving for $\ty$ we get:
 \be \label{hycu2} 
\ty_{\pm}=\frac{(1+\tx+\tx^2
q_s)\pm \sqrt{(1+\tx+\tx^2 q_s)^2-4 q_t \tx^2} } {2}\, \ee
 from which we can construct the meromorphic  differential
 \eqref{logth}. The Bergmann kernel can then be computed in terms of the
branch points of the $\tx$-projection using Akemann's formula \eqref{akemann}. The branch points are given by:
 \be
  \label{bpp1p1} \lambda_{1,2}=
{1\over 2} - {\sqrt {q_t}} \mp {1\over 2} {\sqrt {(1+ 2  {\sqrt
{q_t}})^2 - 4q_s}}, \quad \lambda_{3,4}= {1\over 2} + {\sqrt {q_t}}\pm
{1\over 2} {\sqrt {(1-2  {\sqrt {q_t}})^2 - 4q_s}}. \ee

The  large radius open flat coordinate  for  outer branes is 
given by the integral 
$U=\int_{\alpha_u}\lambda$, where 
the cycle $\alpha_u$ is analogous to the one in figure 2. This 
is evaluated to
\begin{equation}
U=\tilde u-\frac{t_s-T_s}{2}\ ,
\label{p1p1mirrormap}
\end{equation}  
where $T_s$ is the closed flat coordinate.

We now have to implement the phase transition from large radius to the orbifold point. That is, we need to extract the disk and annulus amplitudes at the orbifold point from the large radius ones, as explained in section \ref{movingBmodel}. The disk transforms trivially, hence we just need to expand it in
the appropriate variables at the orbifold point. However, the Bergmann kernel
undergoes a non-trivial modular transformation. 

The phase transition from large radius to orbifold in the local $\IP^1 \times \IP^1$ geometry is given by an $S$-duality transformation of the periods, corresponding to an exchange of the vanishing cycles.\footnote{In fact, this is not quite right. Going from large radius to the orbifold patch not only exchanges the cycles, but also changes the symplectic pairing by an overall factor of $2$. Hence, the transformation is not quite symplectic; this is analogous to the transformation from large radius to the orbifold of local $\IP^2$ considered in \cite{abk}. As was explained there, this change in the symplectic pairing can be taken into account by renormalizing the string coupling constant. In the present case, we get that $g_s = 2 \hat g_s$, where $\hat g_s$ is the Chern-Simons coupling constant. This is also the origin of the $1/2$ factors in \eqref{csvar}.} This is precisely the case that was studied in section \ref{movingBmodel}. 
 This transformation can  be implemented directly into the Bergmann kernel by permuting the branch points
   \be
\label{permute} (\lambda_1,\lambda_2,\lambda_3,\lambda_4)\rightarrow (\lambda_1,\lambda_4,\lambda_3,\lambda_2)\, 
\ee
in Akemann's formula \eqref{akemann}.

All the other orbifold open amplitudes can then be generated by simply
using the new Bergmann kernel (with the new ordering of the cuts) in
the recursion.

\subsubsection{Orbifold flat coordinates}

We will now introduce the orbifold flat coordinates.
Let us start with the closed ones.
 The appropriate  variables to study the orbifold expansion were introduced in \cite{akmvmm}
 and read:
  \begin{eqnarray}
\label{orbivar} q_1=1-\frac{q_t}{q_s}, \   \   \   \  \
q_2=\frac{1}{\sqrt{q_s}(1-\frac{q_t}{q_s}) }.
\end{eqnarray} 
In order to have $q_1$ and $q_2$ both small at the orbifold point,
we have to take the following double---scaling limit:
 \begin{eqnarray}
q_t\rightarrow q_s, \ \ \ \ \ \ \sqrt{q_s}\rightarrow \infty,\ \ \ \ \ \ \ \sqrt{q_s} (1-\frac{q_t}{q_s})\rightarrow \infty, 
\end{eqnarray}
corresponding to a 
blow up in the $(q_s,q_t)$-plane, which was described in 
detail in \cite{akmvmm}.

The  flat coordinates $s_1$ and $s_2$,
 are solutions of the Picard-Fuchs 
equations with a convergent local expansion in the 
variables   $q_1$ and $q_2$.
The principal
structure of the solutions of the orbifold Picard-Fuchs 
equations is 
\begin{align}
\omega_0 &=1, \notag\\
\ \ s_1 &= -\log(1-q_1), \notag\\ 
\ \ s_2 &=\sum_{m,n} c_{m,n} q_1^m q_2^n, \notag\\
\ \ 
F^0_{s_2}&=s_2 \log(q_1)+\sum_{m,n} d_{m,n} q_1^m q_2^n\, ,
\label{closedorbsol}
\end{align}
where the recursions  of the $c_{m,n}$ and $d_{m,n}$ follow from the 
Picard-Fuchs operator.
 Note that the expansion 
coefficients $c_{n,m}$ have the property $c_{m,n\ {\rm mod} \ 2 }=0$. 

The closed flat coordinates are related to the 
't Hooft parameters of Chern-Simons theory. 
The precise relation was found in~\cite{akmvmm} to be
\begin{eqnarray}
\label{csvar} 
S_1=\frac{1}{2} T_1=\frac{(s_1+s_2)}{4},\   \  \   \  \
S_2=\frac{1}{2} T_2=\frac{(s_1-s_2)}{4}\, .
\end{eqnarray}

According to the location of the orbifold divisor at $q_s\rightarrow \infty$, 
described above, $q_2$ picks up a phase under the orbifold monodromy 
$M_{\mathbb{Z}_2}$ around it. 
Therefore, by definition (\ref{csvar}) has the following behavior under orbifold monodromy: 
\begin{equation} 
M_{\mathbb{Z}_2}: (S_1, S_2) \mapsto (S_2, S_1) .
\label{z2action1}
\end{equation}
Notice that the closed string orbifold amplitudes,  calculated in  \cite{akmvmm}, 
are indeed invariant under the above $M_{\mathbb{Z}_2}$ momodromy,  
as required  for an orbifold expansion (see also \cite{abk}). 

Using the explicit form for the periods and the 
relation with the Chern-Simons variables we find the inverse mirror map 
for the closed parameters:
\begin{align}
q_1 &=2
(S_1+S_2)-2(S_1+S_2)^2+\frac{4}{3}(S_1+S_2)^3+\cdots
\notag\\
 q_2 &= \frac{S_1-S_2}{S_1+S_2}+\frac{1}{2}(S_1-S_2)+\frac{(S_1^2 S_2-S_1 S_2^2)}{12
 (S_1+S_2)}+\frac{(S_2^3 -S_1^3)}{24}+\cdots
\label{orbimaps}
\end{align}
We see from this expansion that, as already mentioned, $q_2$ picks up a phase under orbifold monodromy. More precisely, we get the behavior:
\be
M_{\IZ_2}: (q_1, q_2) \mapsto (q_1, - q_2).
\label{monq1q2}
\ee

Let us now consider  the open flat coordinate.
Recalling section \ref{phaseflat}, the open flat coordinate should be a 
linear combination of
\begin{equation}
u_B=\tilde u-\frac{t_s}{2},
\end{equation}
the disk amplitude
\be
A^{(0)}_1(\tx; q_s,q_t)=\int_{\beta_u}\lambda,
\ee
which according to (\ref{globalopen}) 
are  globally defined
integrals, and the closed string 
solutions (\ref{closedorbsol}).
In this case, we fix the six coefficients in the definition of the open flat coordinates by matching the disk amplitude at the orbifold with the result from Chern-Simons theory.



Defining as usual exponentiated coordinates $X_B = \re^{u_B}$ and $\tx = \re^{\tilde u}$, we get for the open flat coordinate $p:= X_{orb} = \re^{u_{orb} }$:
\begin{equation}
\label{p1p1orbcoord}
p:=X_{orb}=X_B = \tx \sqrt{q_s} = { \tx \over q_1 q_2}.
\end{equation}
Expanding the inverse relation
\be
\tx = X_{orb} q_1 q_2,
\ee
we get the open string inverse mirror map
\be
\tx=X_{orb} \left( 2 (S_1-S_2)-(S_1-S_2)(S_1+S_2)+\frac{1}{3}(S_1-S_2)(S_1+S_2)^2 +\cdots
 \right)\, .
\label{orbimaps2}
\ee

Since $\tilde x$ is a globally defined variable on the curve,
 we see from (\ref{p1p1orbcoord}) and \eqref{monq1q2} that under orbifold monodromy,
\begin{equation} 
M_{\mathbb{Z}_2}:X_{orb} \mapsto -X_{orb} \ .
\label{monXorb}
\end{equation}  
This  monodromy behavior of the open flat coordinate is crucial to ensure monodromy
invariance of the topological sting orbifold amplitudes.
This mechanism  is already visible in the first few terms of (\ref{orbimaps2}); under the orbifold $M_{\IZ_2}$ monodromy, the minus sign coming from $S_1 \leftrightarrow S_2$ cancels out with the minus sign coming from the action \eqref{monXorb} on $X_{orb}$, leaving the mirror map invariant.

Furthermore, one can check that 
adding other periods $s_i$, $F^0_{s_2}$ or the disk amplitude $A^{(0)}_1$ to the definition of the open flat parameter would spoil this invariance
property, so that we can fix the open flat parameter $X_{orb}$ uniquely, up to a scale.

\subsubsection{Results}

 We have now all the ingredients required  to compute  open orbifold amplitudes.
 Let's start with the disk amplitude:
 \be
\ba \label{mmorbidisk}
A^{(0)}_1(p)&= p\biggl\{ 2\,S_1 - S_1^2 + \frac{S_1^3}{3} - \frac{S_1^4}{12} - 2\,S_2 + \frac{S_1^2\,S_2}{2} - \frac{S_1^3\,S_2}{4} + S_2^2 -\\ &- \frac{S_1\,S_2^2}{2} - \frac{S_2^3}{3} + \frac{S_1\,S_2^3}{4} + \frac{S_2^4}{12}+ \cdots\biggr\} \\
&+  p^2\biggl\{ S_1 - 2\,S_1^2 + \frac{5\,S_1^3}{3} -
\frac{11\,S_1^4}{12} + S_2 + S_1^2\,S_2 - \frac{3\,S_1^3\,S_2}{2} -
2\,S_2^2 +\\&+ S_1\,S_2^2 - \frac{S_1^2\,S_2^2}{2} +
\frac{5\,S_2^3}{3} -
  \frac{3\,S_1\,S_2^3}{2} - \frac{11\,S_2^4}{12}+ \cdots \biggr\} \\
&+p^3 \biggl\{ \frac{2\,S_1}{3} - 3\,S_1^2 + 5\,S_1^3 -
\frac{59\,S_1^4}{12} - \frac{2\,S_2}{3} + \frac{3\,S_1^2\,S_2}{2} -
\frac{19\,S_1^3\,S_2}{4} + 3\,S_2^2 -\\&- \frac{3\,S_1\,S_2^2}{2} -
5\,S_2^3 +
  \frac{19\,S_1\,S_2^3}{4} + \frac{59\,S_2^4}{12}\cdots \biggr\}
\ea \ee 
Comparing \eqref{mmorbidisk}  with the Chern-Simons result
\eqref{orbidisk},
 we see that to match the two results  we have to
 multiply \eqref{mmorbidisk}  by $-\frac{1}{2}$ and
 send $S_1\rightarrow -S_1$, $S_2\rightarrow -S_2$.
 
With  the above  identifications we also checked that
 the higher amplitudes, such as the annulus, genus 0, three-hole and genus 1, 
one-hole, reproduce the Chern-Simons results. We notice that, as required,
all the higher amplitudes are invariant under the $ M_{\mathbb{Z}_2}$ monodromy.


Framing can also be  taken into account;
 let us see how it goes for the disk amplitude. Higher amplitudes 
 can be dealt with in a similar fashion.
  We start by computing  the
reparameterization $\tx=\tx( x)$ corresponding to the symplectic
transformation
 \be 
(\tx,\ty)\rightarrow (x, y)=(\tx \ty^f,\ty) \, , 
\ee 
which reads:
\begin{eqnarray}
\label{repa} \tx(x)&=&x - f\, x^2 + \frac{\left( f
+ 3\,f^2 - 2\,f\,q_s + 2\,f\,q_t \right) \, x^3}{2} +\dots\,.
\end{eqnarray}
The   bare framed disk amplitude is simply given by:
\begin{eqnarray}
\label{fraord} A_1^{(0)}( x )=\int  \log \ty(\tx(x)) {\rd x \over x}\,.
\end{eqnarray}
We could have computed $y= y( x)$ by solving the framed mirror curve for $y$, rather than by reparameterizing $\ty(\tx)$;
the reparameterization \eqref{repa} is however required to compute the
framed Bergmann kernel.

 We then have to expand the bare disk amplitude  \eqref{fraord}  in the
orbifold variables \eqref{orbivar}, and express the result in flat
coordinates using the inverse mirror maps \eqref{orbimaps} and \eqref{orbimaps2}. Doing so, we obtain a perfect matching with the Chern-Simons result
\eqref{orbifrdisk} once the identification
\be
 f_{cs}= 2 f
 \ee
 between the Chern-Simons integer $f_{cs}$
and the integer $f$ appearing  in the symplectic transformation  is
taken into account. The matching holds for higher amplitudes with the above identification.

\subsection{The $\mathbb{C}^3/\mathbb{Z}_3$  orbifold}
\label{c3z3orbifold}

We studied in detail the open amplitudes at the local $\IP^1 \times \IP^1$ orbifold point, and checked our results with Chern-Simons theory using large $N$ duality. Here we will make a prediction for the disk amplitude at the local $\IP^2$ orbifold point, which corresponds to the geometric orbifold $\IC^3 / \IZ_3$.

Basically, we use the same principles formulated in section \ref{phaseflat} to determine the flat parameters at the orbifold point, up to a scale factor. This is sufficient to predict the disk amplitude. To go to higher amplitudes, we would also need to understand the modular transformation of the annulus amplitude. We are presently working on that and hope to report on it in the near future.

Recall from section \ref{mirrormapP2} that the chain integral giving the open flat parameter at large radius of local $\IP^2$ is given by
\be
U= \tilde u -\frac{t-T}{3},
\ee
and the invariant combination of integrals is
\begin{equation}
u_B=\tilde u - \frac{t}{3}\ .
\end{equation} 
Since this is globally defined, it provides a basis vector for the flat coordinates at the orbifold point. In terms of exponentiated coordinates $X_B = \re^{u_B}$, $\tx = \re^{\tilde u}$ and $q = \re^{-t}$, we get
\begin{equation}
X_B=\tilde x q^{\frac{1}{3}}= (-)^\frac{1}{3}\frac{\tilde x}{3 \psi}  \ ,
\end{equation} 
where we introduced the variable $\psi$ on the moduli space defined by $q=-\frac{1}{(3 \psi)^3}$, so that the orbifold point is at $q\rightarrow \infty$, or $\psi\rightarrow 0$.
 
To determine the open flat coordinate at the orbifold point, we can form combinations of the closed periods, the chain integral $u_B$ and the disk amplitude $A^{(0)}_1(\tilde x,q)$. But we find that
\begin{equation}
X_{orb}=X_B = (-)^\frac{1}{3}\frac{\tilde x}{3 \psi} 
\label{openflatc3z3}
\end{equation} 
is the only combination which leads to a monodromy invariant orbifold disk amplitude.

Using this open flat parameter, we can write down explicitly the disk amplitude for $\IC^3 / \IZ_3$. In \cite{abk}, the closed flat parameter $\sigma$ at the orbifold point was determined, using the Picard-Fuchs equations. We refer the reader to \cite{abk} for the explicit form of $\sigma$ as an expansion in $\psi$ around $\psi=0$. Using this result and the open flat parameter \eqref{openflatc3z3}, we get the following disk amplitude, up to a scale of $X_{orb}$:
\be
\ba 
\label{z3orbidisk}
A^{(0)}_1&=\left( \sigma+ \frac{\sigma^4}{648} - 
     \frac{29\,\sigma^7}{3674160} + \ldots \right) \,X_{orb}\\
   &+ \left( \frac{-\sigma^2}{2} - \frac{\sigma^5}{648} + 
     \frac{197\,\sigma^8}{29393280} \ldots \right) \,X_{orb}^2 \\
   &+ \left(-1+ \frac{\sigma^3}{3}+ \frac{\sigma^6}{648} -
     \frac{\sigma^9}{181440} \ldots\right) \,X_{orb}^3 +{\cal O}({X_{orb}^4})\ .
\ea
\ee
Notice that under the $\mathbb{Z}_3$ orbifold monodromy, given by
\be
\psi \mapsto \re^{2 \pi i \over 3} \psi,
\ee
we have that
\be
M_{\IZ_3}: (X_{orb}, \sigma) \mapsto (\re^{- {2 \pi i \over 3}} X_{orb}, \re^{2 \pi i \over 3} \sigma),
\ee
which leaves the disk amplitude (\ref{z3orbidisk}) invariant, as it should.

\section{Conclusion and future directions}

The formalism proposed in this paper opens the way for various avenues of research. Let us mention a few specific ideas.

\begin{itemize}
\item In this paper we proposed a complete B-model formalism to compute open and closed topological string amplitudes on local Calabi-Yau threefolds. An obvious question is whether we can extend this formalism to compact Calabi-Yau threefolds. At first sight this seems like a difficult task, since we relied heavily on the appearance of the mirror curve in the B-model geometry to implement the recursive formalism of Eynard and Orantin. However, there are various approaches that one could pursue. One could try to generalize the geometric formalism to higher-dimensional manifolds so that it applies directly to compact Calabi-Yau threefolds. Another idea, perhaps more promising, would be to formulate the recursion relations entirely in terms of physical objects in B-model topological string theory; in such a formalism it would not matter whether the target space is compact or non-compact.
\item
We checked our formalism for all kinds of geometries, and have a rather clear understanding of the origin of the recursive solution based on the chiral boson interpretation of the B-model (see \cite{mm2}). It was also proved in \cite{emo} that once the Bergmann kernel is promoted to a non-holomorphic, modular object, the amplitudes that we compute satisfy the usual holomorphic anomaly equations.  But we do not have a proof that our formalism really \emph{is} B-model topological string theory, not even a ``physics proof". It would be very interesting to produce such a proof, probably along the lines of \cite{mm2}.
\item The recursion relations that we used were first found when the curve is the spectral curve of a matrix model. In the local geometries considered in this paper, there is no known matrix model corresponding to the mirror curves. Nevertheless, the recursion relations compute the topological string amplitudes. It would be fascinating to try to find a matrix model governing topological string theory on these local geometries. This could also provide a new approach towards a non-perturbative formulation of topological string theory.
\item Our formalism can be used to study phase transitions in the open/closed moduli spaces, and generate open and closed amplitudes at any point in the moduli space, including in non-geometric phases. We used this approach to study $S$-duality transformations and the orbifold point in the local $\IP^1 \times \IP^1$ moduli space, and compared our results with Chern-Simons expectation values. We also proposed a prediction for the disk amplitude of $\IC^3 / \IZ_3$, which corresponds to the orbifold point in the local $\IP^2$ moduli space. However, while closed orbifold Gromov-Witten invariants are well understood mathematically, to our knowledge open orbifold Gromov-Witten invariants have not been defined mathematically. Hence, it would be fascinating to extend our analysis further and obtain a physics prediction for the higher open invariants of $\IC^3 / \IZ_3$. In order to obtain these results, we would need to understand the Bergmann kernel at the orbifold point; this is more complicated than the $\IP^1 \times \IP^1$ example studied in this paper since the tranformation from large radius to the orbifold is now in $SL(2,\IC)$ --- see \cite{abk}. We are presently working on that and should report on it in the near future.
\item
Notice that the closed and open string amplitudes on $X_p$ provide the 't Hooft resummation at strong coupling of the perturbative amplitudes
of Chern--Simons gauge theory on $L(p,1)$, which is a nontrivial problem for $p>1$. Already in the
simple case of Chern--Simons theory on $L(2,1)$, the resummation problem involves considering a nontrivial moduli space, namely
the moduli space of
complex structures for the mirror of $X_p$, where the orbifold point corresponds to weak 't Hooft coupling and the large radius point corresponds to strong
't Hooft coupling. It would be interesting to see if the lessons extracted from this example
have consequences for the problem of the 't Hooft resummation of $\CN=4$ SYM amplitudes, where a lot
of progress has been made recently. At the very least, the topological example we have solved shows that the analytic structure of the
't Hooft moduli space is very complicated, and that a clever parametrization of this space (by using an analogue of the mirror map)
might simplify considerably the structure of the amplitudes.
\end{itemize}

\appendix

\section{Useful conventions}

In this appendix we remind some useful conventions necessary in order to compare the topological open string amplitudes (\ref{opena}) with the results of the topological vertex. 
In the formalism of the topological vertex \cite{topvertex}, open string amplitudes are encoded in a generating functional depending on a $U(\infty)$ matrix $V$ 
\be
\label{totalopenf}
F(V)=\log \, Z(V), 
\ee
where 
\be
\label{totalopen}
Z(V)=\sum_R Z_R \, \tr_R\, V
\ee
is written as a sum over partitions $R$. It is often convenient to write the free energy $F(V)$ in terms of connected amplitudes in the basis 
labeled by vectors with nonnegative entries $\vec k=(k_1, k_2, \cdots)$. In this basis, 
\be
F(V)=\sum_{\vec k}{1 \over z_{\vec k}!} W^{(c)}_{\vec k} \Upsilon_{\vec k}(V) 
\label{convev}
\end{equation}
where (see for example \cite{mmrev} for details)
\be
\Upsilon_{\vec k}(V) =\prod_{j=1}^{\infty} ({\rm Tr} V^j)^{k_j}, 
\qquad z_{\vec k}=\prod_j k_j! j^{k_j}.
\ee
The functional (\ref{totalopen}) is related to the generating functions (\ref{opena}) as
\be
F(V)=\sum_{g=0}^{\infty}\sum_{h=1}^{\infty} g_s^{2g-2+h} A_h^{(g)} (z_1, \cdots, z_h), 
\ee
after identifying 
\be
\label{idenwords}
 {\rm Tr}\, V^{w_1} \cdots \tr \, V^{w_h} \leftrightarrow m_w (z)=\sum_{\sigma \in S_h} \prod_{i=1}^h z_{\sigma(i)}^{w_i}
 \ee
where $m_w(z)$ is the monomial symmetric polynomial in the $z_i$ and $S_h$ is the symmetric group of $h$ elements. Under this dictionary we have that
\be
 A_h^{(g)} (z_1, \cdots, z_h)=\sum_{\vec k~|~ |\vec k|=h}{1 \over z_{\vec k}!} W^{(c)}_{\vec k}\Upsilon_{\vec k}(V), 
 \ee
 where 
 \be
 |\vec k| =\sum_j k_j. 
 \ee

\end{document}